\documentclass[preprint,12pt]{elsarticle}

\usepackage{amssymb}
\usepackage{amsmath}
\usepackage{bm, adjustbox, color}
\usepackage{comment}
\usepackage{booktabs}
\usepackage{float}
\usepackage{tabularx}
\usepackage{caption}
\usepackage{hyperref}
\usepackage{amsmath}

\usepackage{siunitx}
\usepackage{array} 
\usepackage[utf8]{inputenc}
\usepackage{amsthm}
\usepackage{graphicx}
\usepackage{subcaption}
\usepackage{array,multirow}
\usepackage{siunitx}
 \usepackage{float}
\definecolor{myOrange}{rgb}{1,0.5,0}
\definecolor{myFuchsia}{rgb}{0.7,0.2,0.9}
\definecolor{myBlue}{rgb}{0.4,0.4,1}
\def\BrightOrange{myOrange}
\def\Fuchsia{myFuchsia}
\def\BrightBlue{myBlue}

\long\def\ss#1{%
\par\noindent
{\color{\BrightOrange}$\bullet$~#1}\par}

\long\def\hf#1{%
\par\noindent
{\color{\Fuchsia}$\bullet$~#1}\par}

\long\def\note#1{%
\par\noindent
{\color{\BrightBlue}$\bullet$~#1}\par}

\journal{Mechanics and Physics of Solids}

\begin{document}

\begin{frontmatter}



\title{Characterizing and modeling the mechanical behavior of an anion exchange membrane for carbon capture applications}


\author[inst1]{Sara Sarbaz}
\author[inst1]{Zhi Xin Liu}
\author[inst1]{Heidi Feigenbaum}
\author[inst1]{Samaneh Bayati}
\author[inst2]{Winston Wang}
\author[inst1]{Jennifer Wade}
\author[inst3]{Husain Mithaiwala}
\author[inst3]{Matthew D. Green}


\affiliation[inst1]{organization={Steve Sanghi College of Engineering, Northern Arizona University},
            city={Flagstaff},
            postcode={86011}, 
            state={Arizona},
            country={USA}}
\affiliation[inst2]{organization={ McKetta Department of Chemical Engineering, University of Texas at Austin},
            city={Austin},
            postcode={78712}, 
            state={Texas},
            country={USA}}
\affiliation[inst3]{organization={School for Engineering of Matter, Transport and Energy, Arizona State University},
            city={Tempe},
            postcode={85281}, 
            state={Arizona},
            country={USA}}


\begin{abstract}
A new direct air capture (DAC) technology uses a moisture swing (MS) process with anion exchange membranes, potentially offering a more energy-efficient way to remove CO\(_2\) from the air. In this MS process, the membrane absorbs CO\(_2\) as it dries and releases it when water is added.  Understanding the mechanical behavior of these membranes is essential for improving the design and efficiency of DAC systems and prolonging sorbent lifetime.

This study tested one anion exchange membrane, Fumasep FAA-3, under mechanical loading and various temperature and humidity conditions to measure its swelling, stiffness, strength, plastic deformation, and stress relaxation. Experimental results were used to identify a mechanical model for FAA-3  
that can be used to predict the material’s nonlinear viscous behavior under various loads and environments.

\end{abstract}


\end{frontmatter}



\section{Introduction}

Anion exchange membranes (AEMs) have various uses, such as in fuel cells, batteries, and electrodialysis for water desalination~\cite{SILBERSTEIN20105692}. In this work, we explore a novel application of AEMs: utilizing them for the continuous separation of CO$_2$ from air via a moisture-driven direct air capture (DAC) process.  

DAC removes CO$_2$, the most abundant greenhouse gas, from the atmosphere~\cite{lackner1999carbon}. Given the increasing impact of climate change, its deployment has become essential. DAC, together with safe and permanent CO$_2$ sequestration, is considered a promising approach to achieve net negative greenhouse gas emissions.  
However, most DAC technologies remain energy-intensive and expensive~\cite{DAC}. Using AEMs in moisture-driven DAC may enable continuous membrane-based separation that is more energy-efficient and cost-effective~\cite{castro2022new}.

The AEMs used in DAC contain anion exchange sites that can capture and release CO$_2$ through a moisture swing (MS) process~\cite{MoistureSwingSorbent,kaneko2022kinetic,Wade}.  
In dry conditions, carbonate or hydroxide anions in AEMs chemically bind with CO$_2$ to form bicarbonate anions. When hydrated by liquid water or vapor, these bicarbonate anions release CO$_2$, regenerating carbonate or hydroxide anions (e.g., in materials like IRA900)~\cite{MoistureSwingSorbent}.

Anion exchange materials are often brittle, especially in a partially hydrated state, which may limit the lifespan of MS DAC systems and make their widespread use cost-prohibitive. To better understand their mechanical performance and prevent failure in MS DAC applications, we aim to characterize and model the mechanical behavior of a commercially available moisture-swing anion exchange membrane, Fumasep's FAA-3.  
FAA-3 is based on a lightly cross-linked poly(phenylene oxide) (PPO) backbone and is functionalized with quaternary ammonium groups, charge-balanced by bromide ions~\cite{MOHANTY202027346}.  
The bromide form of FAA-3 is ion-exchanged into an alkaline state (e.g., CO$_3^{2-}$ or OH$^-$) to make it reactive with CO$_2$. As a polymer, this material is expected to undergo permanent deformation and relaxation or creep, making mechanical modeling nontrivial. Specifically, this work examines the mechanical behavior of FAA-3 loaded with CO$_2$ in the bicarbonate state (FAA-3-HCO$_3$) through experimental testing, and identifies and validates a model for its mechanical response.

The mechanical behavior of FAA-3 and similar anion exchange membranes, including poly(vinyl benzyl chloride)-based membranes (c(Je)-QAPVB) and other PPO-based AEMs, has been previously reported~\cite{Khalid,ZHENG2024854,narducci2016mechanical}.  
However, most of these characterizations are limited to basic stress-strain measurements, with swelling data typically collected only under submerged conditions.

\begin{figure}[hbt!]
    \centering
    \includegraphics[width=0.7\linewidth]{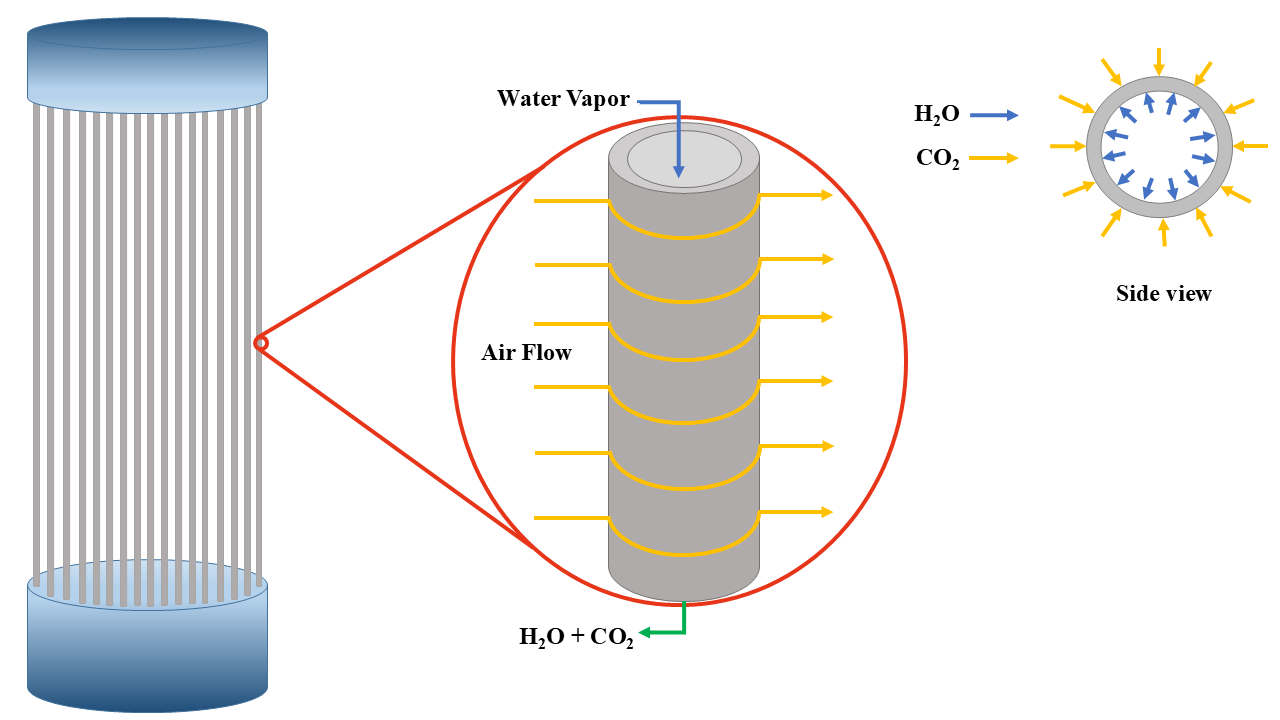}
    \caption{Schematic of the Carbon Capture setup using FAA-3 hollow fibers.}
    \label{fig:schematic}
\end{figure}

Limited data are available on the effects of loading rate, hydration, temperature, or plastic deformation in anion exchange membranes (exceptions include~\cite{SILBERSTEIN20105692,YOON20113933,solasi2008time,krewer2018impact}, though these studies focus on Nafion, a perfluorinated proton exchange membrane). For these reasons, this study examines the influence of these factors in detail for FAA-3 and uses existing AEM data to confirm similar trends and properties, as well as to identify and validate a model for its mechanical behavior.
Since the material is a membrane, the moisture swing process can be made efficient through continuous separation via hollow fibers.  This novel hollow fiber design is shown in Figure~\ref{fig:schematic}.  On the outer surface of the hollow fibers, CO$_2$ is captured from dry arid environments, where the sorbent binds it, while moisture diffuses outward from the inner surface, sustaining the gradient and enabling continuous absorption and desorption. 
This design offers a high surface-area-to-volume ratio, improving mass transfer efficiency and enabling rapid absorption and desorption.
Moreover, this design shows promise for scalable, low-energy CO$_2$ separation for DAC \citep{narducci2016mechanical, ZHENG2024854}. As such, this design motivates the expriments and modeling of FAA-3 in this work.


During moisture update, the material will experience hygroscopic swelling \citep{MOHANTY202027346}, and this swelling can induce expansion on the wet side and/or contraction on the dry side of the hollow fibers, which in turn leads to bending of the hollow fibers.  This bending can generate internal stresses that may lead to micro-cracking, fatigue, or structural failure, particularly near the fixed ends of the hollow fibers where deformation is constrained \citep{narducci2016mechanical}. 
Additionally, airflow around the structure may act like a distributed pressure along the length of the tube, potentially exacerbating the bending due to hygroscopic swell in the upwind direction.


As bending is the primary mode of deformation in this design, our experiments focused on the uniaxial tensile, cyclic, and time-dependent stress-strain response of FAA-3, including hydration and temperature effects. 
Additionally, we characterized the hydration swell and thermal expansion of FAA-3-HCO3 as environmental changes in temperature and moisture are expected. 
Moreover, since a continuous moisture gradient through the thickness of the hollow fibers is part of this design (see Figure \ref{fig:schematic}), the hygroscopic swell was determined through the thickness of the material, even though all other behavior is characterized axially.  
Using these results, we developed a mechanical model incorporating a Neo-Hookean spring for nonlinear elasticity and the Bergström-Boyce model for strain rate-dependent viscoplasticity 
Future work will couple mechanical deformations with CO$_2$ and moisture transport and study the durability and efficiency of the hollow fiber MS DAC system.  
This coupling will elucidate how localized hygroscopic swell from moisture transport leads to mechanical deformation, and how mechanical deformation could alter moisture diffusion rates, impacting CO$_2$ capture efficiency. 
Such a chemo-mechanical analysis will be essential to guide the selection of materials, fiber geometries, and operating conditions to optimize the long-term performance  MS DAC systems.

Thus, this work lays the foundation for designing robust, efficient DAC systems capable of meeting future decarbonization demands.

\section{Materials and Methods}
\label{Materials and methods}
This paper examines the mechanical behavior of a moisture-swing (MS) responsive anion exchange membrane (AEM) using dynamic mechanical analysis (DMA), ultimate tensile testing, load-hold-unload testing, and cyclic loading experiments (Sections~\ref{sec:Dynamic Mechanical Analysis Test} to \ref{sec:Cyclic Loading Test}).  
All tests were conducted in uniaxial tension, with the loading direction oriented axially (in-plane), unless otherwise specified.

DMA was used to evaluate the extent of viscous mechanical behavior, determine the storage and loss moduli, and identify the glass transition temperature.  
Ultimate tensile tests characterized the material’s strength, elastic modulus, and elongation at failure under varying strain rates and humidity levels.  
Load-hold-unload tests assessed the material's viscous response during the hold period and quantified the plastic deformation remaining after unloading.  
Cyclic tests simulated operating conditions that may occur in practical applications, such as changes in airflow or fluctuations in the moisture gradient due to system start-up or shut-down. These tests also served to validate the mechanical models identified in this study.

In addition, tests without any mechanical loading were conducted to characterize the polymer’s hygroscopic swelling strain as a function of water concentration (Section~\ref{sec:Hygroscopic Swell}) and thermal strain as a function of temperature (Section~\ref{sec:Temperature Test}).

The total engineering strain of the material is assumed to be the sum of mechanical, thermal, and hygroscopic contributions, expressed as:
\begin{equation}
    \label{Eq Strain}
    \varepsilon^{\text{tot}} = \varepsilon^{\text{mech}} + \varepsilon^{\text{th}} + \varepsilon^{\text{hygro}},
\end{equation}
where $\varepsilon^{\text{tot}}$ is the total engineering strain, $\varepsilon^{\text{mech}}$ is the mechanical strain, $\varepsilon^{\text{th}}$ is the thermal strain, and $\varepsilon^{\text{hygro}}$ is the hygroscopic swelling strain.  
Each component must be independently characterized to enable accurate mechanical modeling of the material.

Most tests were performed on at least two samples to ensure repeatability. Details of the material and each experimental procedure are described in the following sections.

\subsection{Materials}
\label{sec:Materials}
FAA-3 membranes (30~$\mu$m thick), commercially available from Fumasep, were purchased from the FuelCell Store~\cite{fumasep_faa3}. The membranes were received in the bromide counterion form. To convert the counterions, the samples were ion-exchanged into the bicarbonate form using two 24-hour immersions in 0.1~M KHCO$_3$ solution, separated by a rinse in deionized water.  

\begin{figure}[H]
        \centering
        \includegraphics[width=0.65\linewidth, height=0.3\textheight]{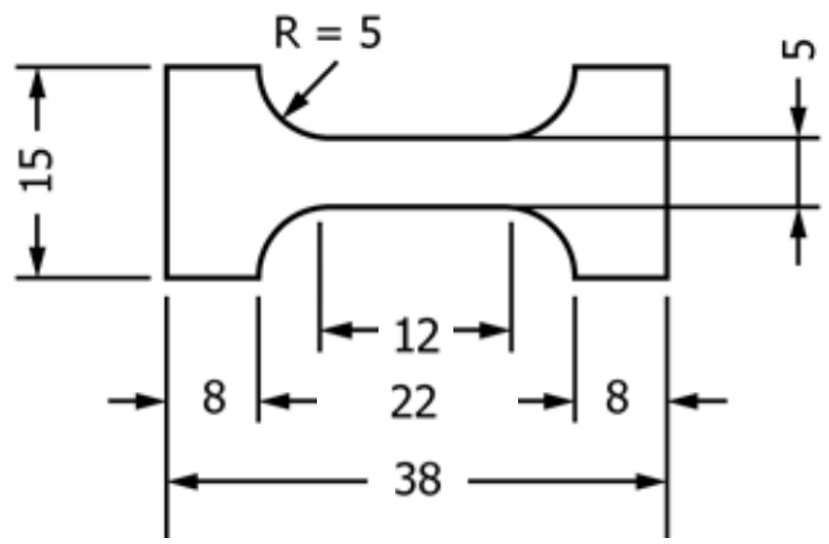}
        \caption{Schematic diagram of the ASTM D-1708, dog-bone sample geometry in millimeters adapted from \cite{ASTM}.}
        \label{fig:ASTM 1708}
    \end{figure}

The bicarbonate-exchanged membranes were equilibrated to ambient conditions—approximately \SI{23}{\celsius} and 30--40\% relative humidity (RH)—prior to being cut into test geometries. Tests were performed using either rectangular specimens or ASTM D1708 dog-bone specimens (Figure~\ref{fig:ASTM 1708})~\cite{ASTM}. The ASTM specimens had a gauge length of 12~mm and a width of 5~mm. Unless otherwise specified, this 12~mm gauge length was used as the reference state for all strain calculations; thus, the reference state corresponds to ambient conditions.

\subsection{Dynamic Mechanical Analysis Tests}
\label{sec:Dynamic Mechanical Analysis Test}

Dynamic mechanical analysis (DMA) is a technique used to characterize the mechanical properties of materials under oscillatory loading. Adwas used to perform DMA testing on FAA-3 membranes. 

The DMA test was performed on a rectangular sample (approximately 7.5~mm $\times$ 12.5~mm under ambient conditions without load). The test involved a temperature sweep from \SI{-150}{\celsius} to \SI{250}{\celsius} at a rate of \SI{3}{\celsius}/min. The sample was cyclically loaded under displacement control at 1~Hz with an amplitude of 18~$\mu$m. This amplitude is similar to that used by Silberstein et al.~\cite{SILBERSTEIN20105692} in their DMA tests on Nafion.  

To reach \SI{-150}{\celsius}, liquid nitrogen was used to cool the sample. Before loading, the chamber was set to the minimum temperature, and the sample was allowed to equilibrate for \SI{20}{\minute}. 


An initial force of 5~N was applied before cyclic loading began, and the force-tracking function was used to maintain a load between 1--5~N to prevent sample buckling. Note that unlike all other tests, the length in the coldest state after applying the 5~N preload was considered the initial length, and this value was 7.76~mm. This preload remained throughout the test. As the temperature increased and the sample expanded, the force never dropped below 1~N, so no additional tensile force was needed during the temperature sweep.

\subsection{Hygroscopic Swelling}
\label{sec:Hygroscopic Swell}

Hygroscopic swelling strain, $\varepsilon^{\mathrm{hygro}}$, from Eq.~(\ref{Eq Strain}), is particularly important for MS DAC applications, as moisture changes are the driving mechanism. The strain $\varepsilon^{\mathrm{hygro}}$ can be related to changes in moisture content by:
\begin{equation}
    \label{Eq hygro}
    \varepsilon^{\mathrm{hygro}} = \bm{\beta} \, \Delta WC,
\end{equation}
where $\bm{\beta}$ is the swell coefficient and $\Delta WC$ is the change in water concentration \cite{Hygroscopic}, in units of mol/m$^3$. When FAA-3 absorbs water, $\Delta WC$ is positive, and $\varepsilon^{\mathrm{hygro}}$ is positive, indicating that the sample will enlarge. When the water evaporates and the sample dries, $\Delta WC$ is negative, and according to Eq.~(\ref{Eq hygro}), that implies $\varepsilon^{\mathrm{hygro}}$ is negative, so the sample will shrink. 

The swell response of the membrane in each direction may be anisotropic, so $\bm{\beta}$ will take the following form:

\begin{equation}
    \label{Eq beta}
    \bm{\beta} = 
    \begin{bmatrix}
        \beta_{\text{in-plane}} & 0 & 0 \\
        0 & \beta_{\text{in-plane}} & 0 \\
        0 & 0 & \beta_{\text{thick}}
    \end{bmatrix}
\end{equation}
where $\beta_{\text{in-plane}}$ is the swell coefficient in-plane (the axial direction), which is expected to differ from $\beta_{\text{thick}}$, the swell coefficient through the thickness (the transverse direction). Preliminary testing showed that the material was transversely isotropic, so it was assumed that the swell coefficient is the same in both in-plane directions.

\begin{figure}[htbp]
        \centering
        \includegraphics[width=0.7\linewidth, height=0.3\textheight]{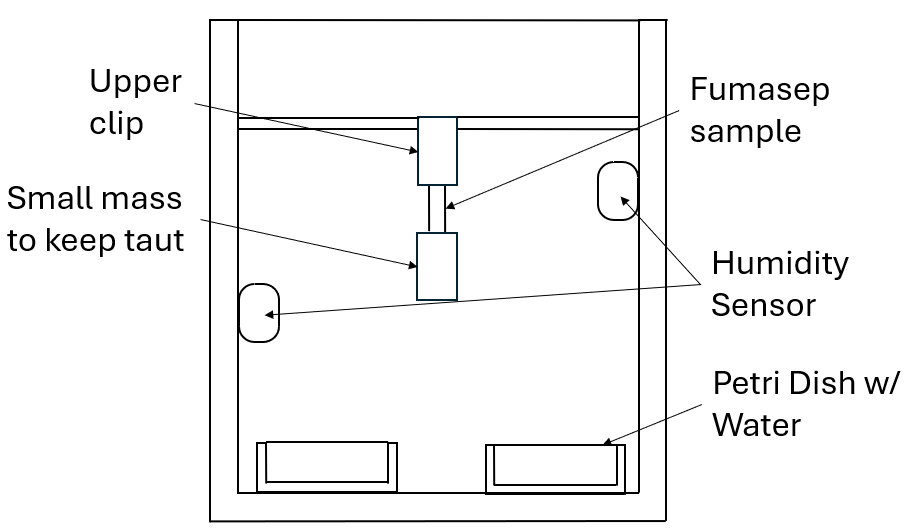}
        \caption{Experimental set-up for the swell test in a humidity chamber.}
        \label{fig:Swell Test Setup}
    \end{figure}

To measure swelling in both the in-plane and thickness directions, a humidity chamber was constructed using plastic wrap (see Figure~\ref{fig:Swell Test Setup}). The sample was suspended vertically within the chamber, and a small mass (8.8~g), considered negligible, was applied to keep the sample taut. Water was placed inside the chamber to humidify the environment, and the opening in the plastic wrap was adjusted to regulate the humidity level around the sample.

The relative humidity within the chamber was varied across four conditions: ambient ($\sim$32\% RH), $\sim$50\% RH, $\sim$80\% RH, and $\sim$90\% RH. The sample was allowed to equilibrate for 24~hours at each humidity level. After equilibration, the length and thickness of the sample were measured to calculate the hygroscopic strain.

The length of the sample is measured with a caliper with an accuracy of $\pm$0.01~mm, and measurements of the thickness were collected using a micrometer, with an accuracy of ${\pm}$0.001~mm. 
Because of the high measurement uncertainty of these tools, 4–6 readings were taken in each direction at each humidity level.
In addition, the chamber opening was often enlarged to allow the calipers and micrometer access to the specimen. This caused temporary fluctuations in humidity during the measurement process, so the humidity at the time of measurement was also recorded.
    
\subsubsection{Water Uptake and Water Content}
During the hygroscopic swelling tests, relative humidity is measured, but water concentration is not.  To use Eq. (\ref{Eq hygro}), we need to relate the measured relative humidity (water activity) to the water concentration (water content) inside the material.  
Towards that end, the water vapor sorption of the membrane was measured using a Discovery SA Dynamic Vapor Sorption Analyzer(TA Instruments, Delaware, USA) with a nitrogen purge rate of 20mL/min. About 1-7 mg of the membrane films were used for each measurement. In this case, 50$\mu$m thick samples were used, but the samples were subject to the same bicarbonate exchange pretreatment. The samples were dried in the instrument at 25$^\circ$C until the sample weight reached equilibrium ($<$0.01 wt\% change in 30 min, 4-12 h). Sample weight at equilibrium is recorded as the dry weight ($m_\text{dry}$). Relative humidity was then increased in 5\% steps from 0\% to 20\% RH, then 10\% steps from 20\% to 90\% RH. For each step, the sample was equilibrated for 8h or until the change in sample weight was less than 0.01\% in 30 min. The water vapor uptake was measured with 3 distinct samples.

To enable meaningful comparisons across membranes with different ion exchange capacities (IEC), water content is often normalized using the dimensionless hydration number, $\lambda$, which represents the number of water molecules per fixed charge site.

Experimentally, water uptake is typically reported as the ratio of grams of sorbed water ($W_{\text{wet}}$) to grams of dry membrane ($W_{\text{dry}}$). Conversion to $\lambda$ is done using the following relation:

\begin{equation} 
\label{Eq lambda}
\lambda = \frac{\frac{W_{\text{wet}} - W_{\text{dry}}}{M_{\text{H}_2\text{O}}}}{\text{IEC} \times W_{\text{dry}}}
\end{equation}
where $M_{\text{H}_2\text{O}} = 18$~g/mol is the molar mass of water, and the IEC (ion exchange capacity) of FAA-3 is 1.85$\times 10^{-3}$~mmol/g~\cite{fumasep_faa3}.
This normalization enables a more consistent interpretation of membrane hydration relative to charge density, providing a direct link between water uptake, swelling behavior, and ionic conductivity.

\subsection{Thermal Expansion}
\label{sec:Temperature Test}
Just as the material swells due to increased water content, temperature changes also lead to normal strains in the material. The thermal strain, $\varepsilon^{th}$, can be related to changes in temperature by 
\begin{equation}
    \label{Eq temp}
    \varepsilon^{th} = \alpha{\Delta}T,
\end{equation}
where $\alpha$ is a thermal expansion coefficient, and ${\Delta}T$ is a change in temperature. Note that Eq. (\ref{Eq temp}) for thermal expansion is analogous to Eq. (\ref{Eq hygro}) for hygroscopic swell. However, in this case, we only measure in-plane thermal expansion due to small changes in ambient temperature. This is because thermal changes are expected to be minimal during the MS process and are primarily caused by changes in the natural environment. 

\begin{figure}[hbt!]
        \centering
        \centering (a){{\includegraphics[width=0.3\linewidth, height=0.24\textheight]{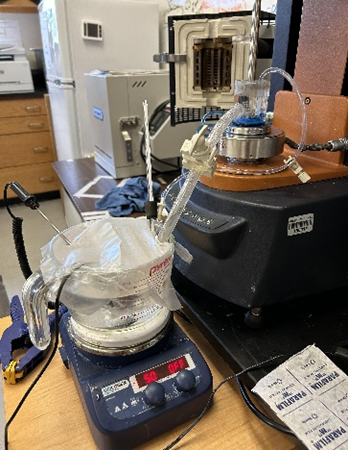}}}%
        \qquad
        \centering (b){{\includegraphics[width=0.4\linewidth, height=0.25\textheight]{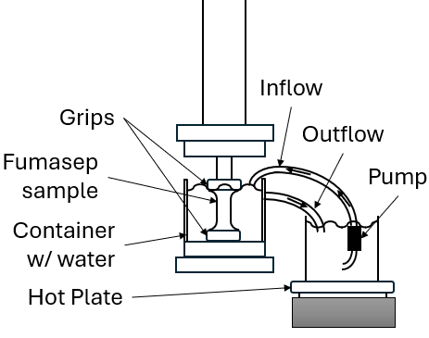}}}%
        \caption{Setup for testing thermal expansion in submerged conditions using an external hotplate to control temperature, (a) picture and (b) schematic.}%
        \label{fig:Rheometer Fluid}%
    \end{figure}

The test was conducted on a TA Instruments Hybrid Discovery Rheometer 2.  The specimen was submerged in DI water, and the axial strain was measured as the temperature of the water increased.  For the duration of the test, the force was set to near zero (a small force of 0.1--0.4~N kept the specimen taut).
Submerging the samples in DI water ensures 100\% RH for all tests (heated air may have resulted in different humidities at different temperatures). To get a fast response of the fluid to changes in temperature, an external system was employed to heat the water. Figure \ref{fig:Rheometer Fluid} shows the hot plate used to heat the water needed for the test and the pump for circulating the fluid into and out of the container enclosing the sample. Before thermal loading, the sample was given \SI{15}{\minute} to equilibrate and swell in water.  The temperature began at ambient room temperature (\SI{21}{\celsius}) and was increased at a rate of approximately \SI{1}{\celsius}/min until the water bath reached a temperature of reached of \SI{40}{\celsius}.
Note that $\alpha$ generally changes, albeit smoothly, around the glass transition temperature, but in the temperature ranges we are testing, it is expected to be constant.  

\subsection{Ultimate Tensile Test}
\label{sec:Ultimate Tensile Test}

Ultimate tensile tests were conducted to characterize the mechanical behavior of the polymer, including ultimate strength, elongation at failure, and elastic modulus. 

An Instron ElectroPuls E3000 system was used to control the strain rate applied to the sample. According to the manufacturer’s specifications, the system’s force resolution is the greater of 0.5\% of the reading or 0.005\% of the load cell capacity (5~kN in this case).

During these and all subsequent mechanical tests conducted in air, ambient temperature and relative humidity were monitored using a hygrometer-thermometer. Environmental conditions remained stable throughout the tests, with minimal fluctuations.

Prior to testing, a nominal strain of approximately 0.2\% was applied to ensure the sample was taut. The sample was then loaded in tension at the specified strain rate until fracture.

\subsubsection{Rate Variance}
\label{sec:Ultimate Tensile Test Rate}
The tensile tests were performed under constant engineering strain rates of 0.1~s$^{-1}$ (fast) and 0.01~s$^{-1}$ (slow), at room temperature and ambient relative humidity.

\subsubsection{Humidity Variance}
\label{sec:Ultimate Tensile Test Humidity}
\begin{figure}[hbt!]
        \centering
        \centering (a){{\includegraphics[width=0.3\linewidth, height=0.24\textheight]{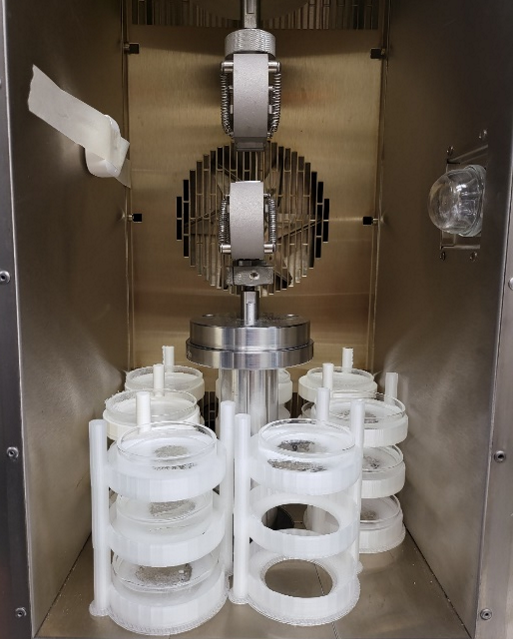}}}%
        \qquad
        \centering (b){{\includegraphics[width=0.45\linewidth, height=0.25\textheight]{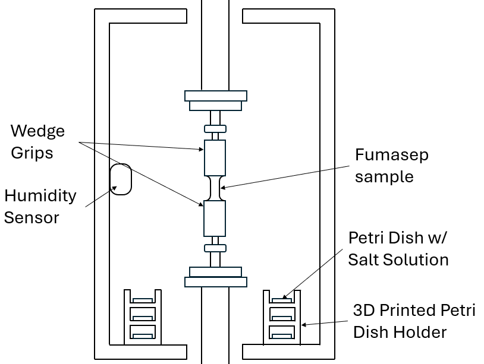}}}%
        \caption{Experimental set-up for the humidity varied ultimate tensile tests in a sealed chamber around the ELECTROPULS E3000 Instron with salt solutions to control humidity, (a) picture and (b) schematic.}%
        \label{fig:Instron Setup}%
    \end{figure}

The ultimate tensile test with varying humidity was performed under a constant engineering strain rate of 0.01 s$^{-1}$ and at room temperature. To control the humidity, the Instron is equipped with a chamber that encloses the sample. Trays of potassium nitrate salt solutions were used to control chamber humidity, as shown in Figure \ref{fig:Instron Setup}. 

Once at equilibrium, the relative humidity was measured with a humidity sensor and recorded. Samples were equilibrated to the humidified chamber RH for $\sim$30~h, prior to testing.

Total strain was measured, and hydroscopic strain was subtracted from this to isolate the mechanical strain. 
Assuming that the material has gauge length $L_0$, length $L_i$ after equilibrating to the humidified environment, and length $L_f$ at any instant during the test, then the mechanical strain can be found by:
\begin{equation}
    \label{Eq mech}
    \varepsilon^{\text{mech}} = \varepsilon^{\text{tot}} - \varepsilon^{\text{hygro}} = \frac{L_f - L_0}{L_0} - \frac{L_i - L_0}{L_0} = \frac{L_f - L_i}{L_0} = \frac{\Delta L}{L_0}
\end{equation}
where ${\Delta}L$ represents the change in length measured during the test relative to the humidified length (i.e. ${\Delta}L = L_f - L_i$).  

\subsection{Load-Hold-Unload Test}
\label{sec:Load-Hold-Unload Test}
The load-hold-unload test was designed to observe the inelastic mechanical properties of FAA-3. Stress relaxation can be observed as the force decreasing while the displacement is fixed.  Plastic deformation is observed as residual strain when unloaded to $\sim$0~N force, as a reminder, to ensure the membrane remains taut, stress is never fully brought to zero.  It is important to note that because the unload portion of the test is not held, it is unclear if the plastic deformation seen in the experiments is truly permanent, or just residual strain that will be recovered with additional time.  

\subsubsection{Humidity Variance}
\label{sec:Load-Hold-Unload Test Humidity}
For the load-hold-unload experiment, the sample is subjected to the following: 1) loaded at the specified strain rate until the sample reaches a mechanical engineering strain of 1\%, 2) held in place for 1 min, 3) unloaded to $\sim$0~N at the same strain rate, 4) reloaded to a mechanical engineering strain of 5\%, 5) held in place for 1 min, 6) unloaded to a $\sim$0~N, 7) reloaded to a mechanical engineering strain of 8\%, 8) hold in place for 1 min, and 9) unloaded to a force of zero.   
These tests were performed using an Instron 8874 machine, with a 50~N $\pm$0.25\% load cell. 

To observe how the mechanics of the material changes with humidity, the tests were conducted at ambient conditions (35\% RH -- 40\% RH), $\sim$70\% RH, and $\sim$95\% RH. 
Prior to testing, the samples were given 24~h in the humidity controlled environment to reach equilibrium. 
The tests were performed under a constant engineering strain rate of 0.01~s$^{-1}$ and at room temperature (~\SIrange{23}{27}{\celsius}).


    \begin{figure}[hbt!]
        \centering
        \centering (a){{\includegraphics[width=0.3\linewidth]{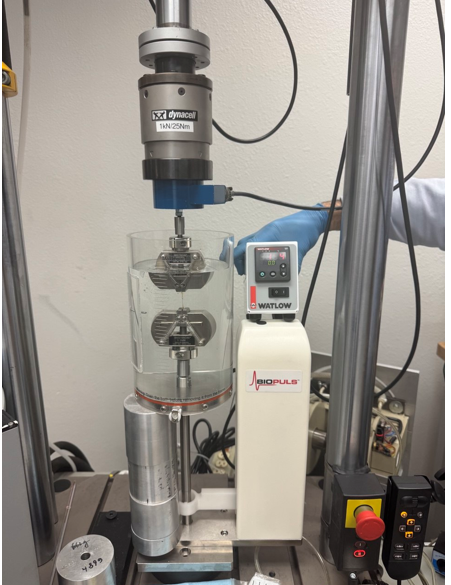}}}
        \qquad
        \centering (b){{\includegraphics[width=0.4\linewidth]{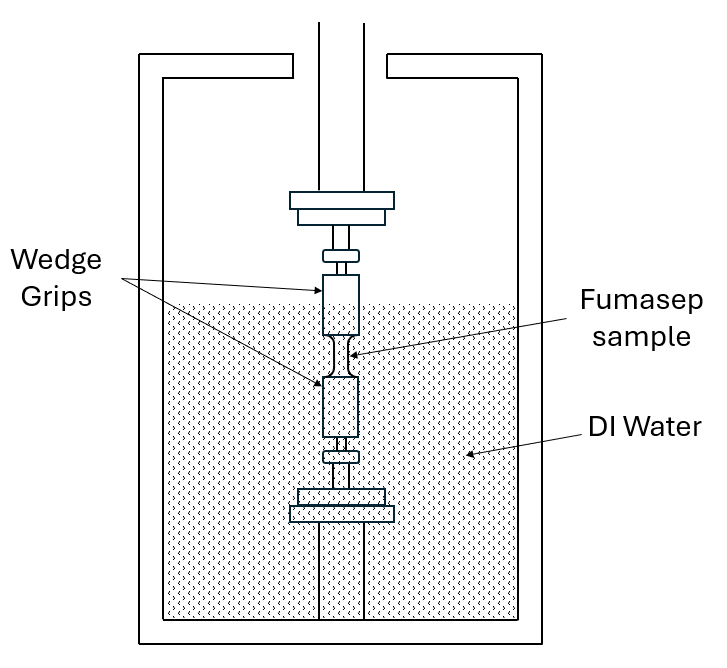}}}
        \caption{Experimental set-up for the temperature varied load-hold-unload tests in using the Instron 8874 and fluid chamber to submerge the sample, (a) picture and (b) schematic.}%
        \label{fig:Fluid Instron Setup}%
    \end{figure}

\subsubsection{Temperature Variance}
\label{sec:Load-Hold-Unload Test Temperature}

The load-hold-unload tests with temperature variance were performed again on the Instron 8874 machine with the 50~N load cell at a constant engineering strain rate of 0.01~s$^{-1}$ and at 100\%RH (submerged). The loading procedure in this case was: 1) load to a mechanical engineering strain of 1\%, 2) hold in place for 1 min, 3) unload to $\sim$0~N, 4) reload to a mechanical engineering strain of 5\%, 5) held in place for 1 min, 6) unload to $\sim$0~N, 7) reload to a mechanical engineering strain of 8\%, 8) hold in place for 1 min, and 9) unload to a force of zero.

To investigate the material's behavior under varying temperatures, the test was conducted at \SI{28}{\celsius} and \SI{40}{\celsius}. 
These values were chosen to mimic ambient changes in temperature found in warm arid climates.  
Before testing, each sample was soaked in water for at least  15 ${\min}$ to allow for swelling and equilibration to temperature.

Similar to Eq. (\ref{Eq mech}), the mechanical strain was found by subtracting the thermal and hygroscopic strains from the total strain, \begin{equation}
    \label{Eq mech temp}
    \varepsilon^{mech} = \varepsilon^{tot}-\varepsilon^{th}-\varepsilon^{hygro} =\frac{{\Delta}L}{L_0}.
\end{equation}
where changes in length are relative to the thermally loaded and swelled state (i.e. ${\Delta}L = L_f - L_i$), and the initial length, $L_0$, is the gauge length before any thermal/hydro/mechanical loading.  

\subsection{Cyclic Loading Test}
\label{sec:Cyclic Loading Test}
The cyclic loading test was designed to simulate natural cycles in humidity and temperature that might be seen during DAC and to verify the model developed for the material. While experimental limitations prevent the change of moisture during a test, we approximated moisture changes by subjecting the material to elongation/release cycles to mimic shrink/swell cycles that may be seen during start-up/shut-down cycles using hollow fibers for MS DAC. Since the material takes time to absorb/desorb water during MS DAC, the cyclic loading test was performed at a strain rate of 0.0001~s$^{-1}$. This test was performed at room temperature and ambient relative humidity (approximately 25--28~${^\circ}$C
 and 37--41\% RH) on two samples.

In these tests, the sample was loaded and unloaded for three cycles. The membrane is subject to the following: 1) loaded at the specified strain rate until an engineering strain of 8\%, 2) released until it reaches $\sim$0~N, 3) reloaded back to an engineering strain of 8\%, 4) unloaded to $\sim$0~N, 5) reloaded back to an engineering strain of 8\%, and 6) unloaded to a force of 0~N.

\section{Experimental Results}
\label{sec:exps}
This section presents the results from the experimental tests used to characterize the mechanical behavior of FAA-3. The material’s response to applied loads, along with the influence of moisture and temperature on its behavior, are analyzed. These results can help identify a model that accurately simulates the mechanics of FAA-3.

The results presented use engineering strain and nominal stress. Therefore, the term ``stress'' implies nominal stress, which is calculated using the reference cross-sectional area. For engineering strain, an additive decomposition is assumed using Eq. (\ref{Eq Strain}). The mechanical part of the engineering strain is often used to present results, as the model for the hygroscopic swell is given in Eq. (\ref{Eq hygro}) and the thermal expansion is given in Eq. (\ref{Eq temp}), thus only a model for the mechanical part must be determined. Moreover, the mechanical strain must capture the viscous effects and plastic deformation. The zero stress/strain state is after the material is taut, so while some stress may be present, it is neglected, and all subsequent stress values are stresses above this minimal value to keep the specimen taut.

\subsection{Dynamic Mechanical Analysis Test}
\label{sec:Dynamic Mechanical Analysis Results}
The results of the dynamic mechanical analysis (DMA) test reveal both the elastic and viscous material response of FAA-3 simultaneously through the storage modulus and the loss modulus. 
    \begin{figure}[hbt!]
        \centering
        \includegraphics[width=0.7\linewidth]{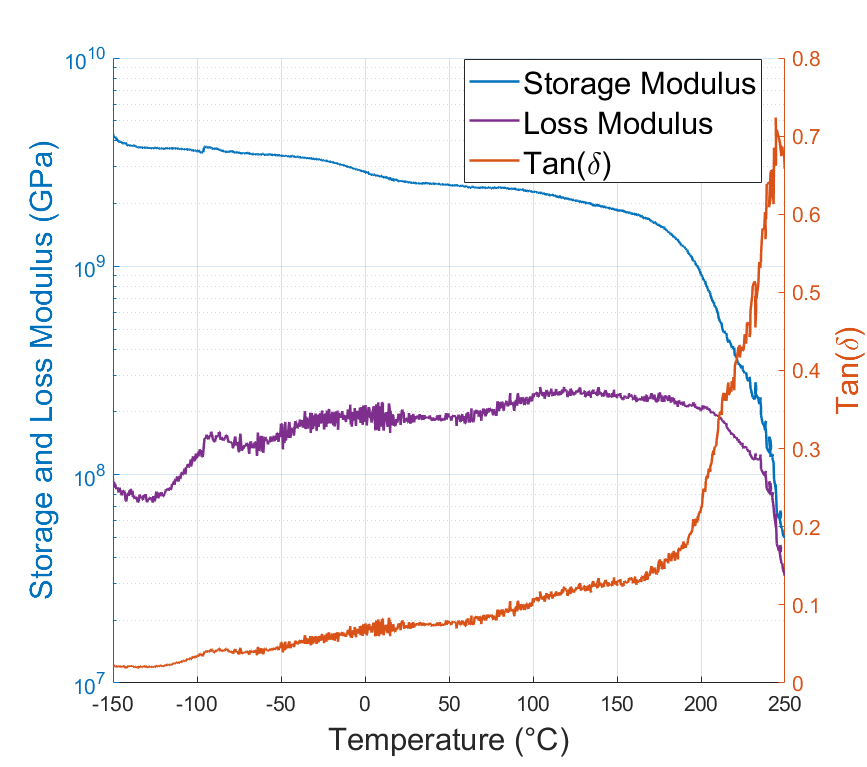}
        \caption{Storage modulus, loss modulus, and damping factor over temperature from DMA testing }
        \label{fig:DMA results}
    \end{figure}

Figure \ref{fig:DMA results} shows that the storage modulus is greater than the loss modulus across the entire temperature range. However, the loss modulus values of 100 MPa are substantial, so viscous effects are prevalent within the material. Figure \ref{fig:DMA results} also shows that the storage and loss moduli are fairly steady throughout the range of temperatures tested, except between \SIrange{200}{250}{\celsius}. When the temperature reached \SIrange{200}{250}{\celsius}, the storage and loss moduli both showed a significant drop. The significant drop in storage modulus and the rise in damping factor ($\tan\delta$), the ratio of the loss modulus to the storage modulus, show that a potential glass transition temperature ($T_g$) may have been reached.

Even though the glass transition temperature could only be inferred to be around \SIrange{200}{250}{\celsius} by this test, the test confirmed that $T_g$ would not likely be in the operational range of temperatures. 
This is important for the applications because when FAA-3 is implemented for direct air capture, the material may be subjected to the temperatures of the surrounding environment (approximately \SIrange{0}{50}{\celsius}), and the mechanical behavior is likely to change around $T_g$. However, it is possible that at higher relative humidity levels, the glass transition temperature may be lower than what is shown in Figure \ref{fig:DMA results}. 

Note that, unlike traditional DMA tests, there is an initial load applied to the sample. Typically, DMA tests only apply a sinusoidal strain. A force of 5~N was applied in this case to keep the specimen taut even as it thermally expands. As the temperature increased, the force on the specimen decreased due to the elongation of the specimen. This may explain the steady decline of the storage modulus before around 200~$^\circ$C, where it might be expected to be flat.  



\subsection{Hygroscopic Swell}
\label{sec:Hygroscopic Swell Results}

\subsubsection{Water Uptake and Water Content}
 \label{water uptake and content}   

\begin{figure}[hbt!]
        \centering
        \includegraphics[width=0.62\linewidth, height=0.35\textheight]{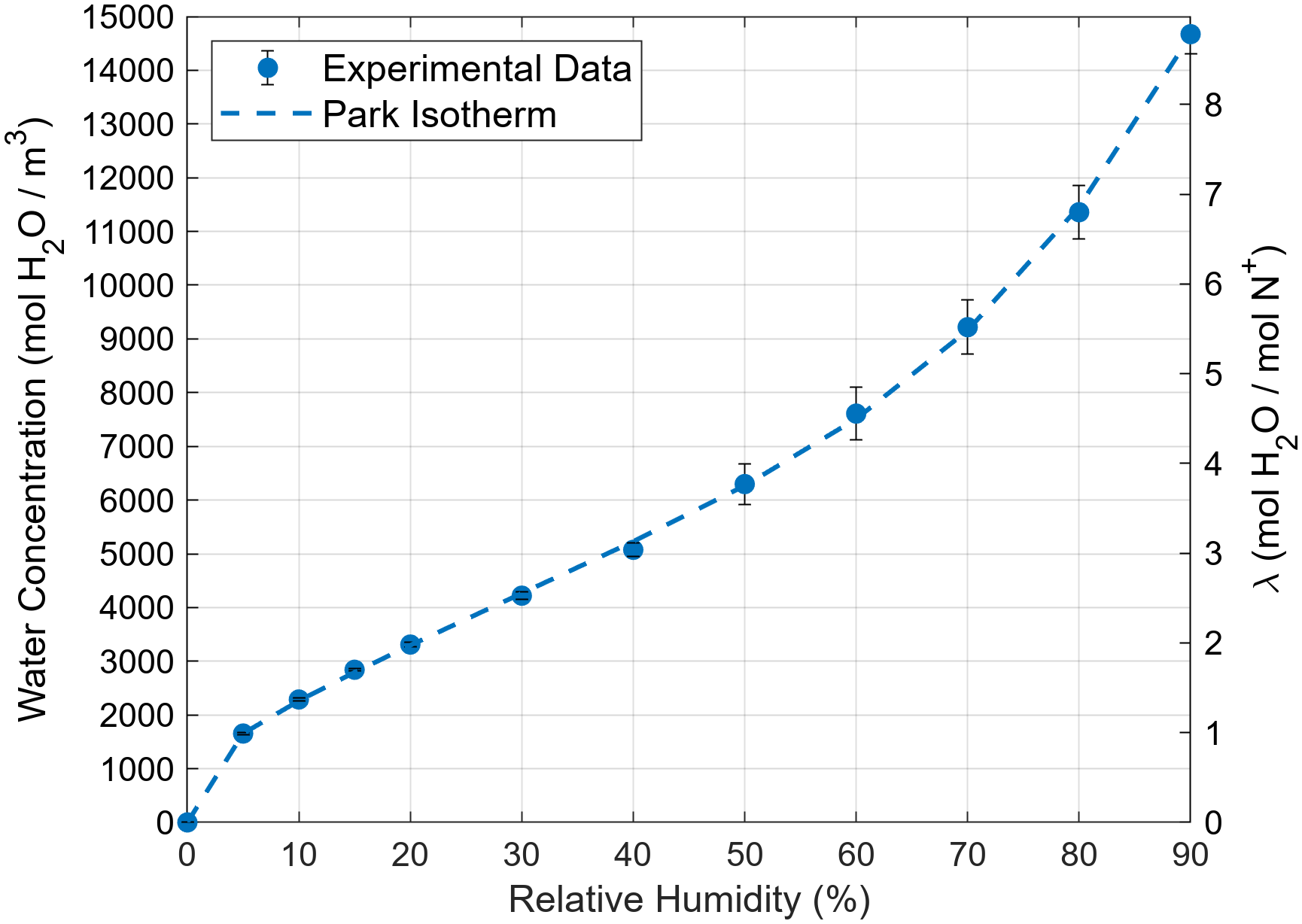}
       \caption{Water sorption data for Fumasep FAA-3 -- HCO\textsubscript{3} measured at \SI{35}{\celsius}, shown alongside a fitted Park model. The data is represented as weight percent (left axis) and as mol H\textsubscript{2}O per mol charge (\(\lambda\), right axis).}
        \label{fig:con vs hum}
    \end{figure}

Water uptake is a critical parameter that influences the mechanical, ionic, and gas transport properties of ion-exchange membranes such as Fumasep FAA-3. 
 Figure~\ref{fig:con vs hum} shows the average water concentration from the three samples
used in the water uptake experiments, with error bars showing the standard deviation.


The water vapor isotherm shown in Figure \ref{fig:con vs hum}
is Type II Sigmoidalin behavior \cite{ThommesKanekoNeimarkOlivierRodriguezReinosoRouquerolSing+2015+1051+1069} and represents water sorption into both micro-voids of the glassy polymer structure and the solution into the dense polymer matrix, which grows exponentially as the water plasticizes the polymer  \cite{park1986transport}.  The behavior was fit using the model developed by Park \cite{park1986transport} to interpolate water loading at a given relative humidity (water activity): 
\begin{equation}
    \label{Eq water concentration}
     W_{\mathrm{H_2O}}=\frac{(A_L b_L a_w)}{(1+b_L a_w )}+k_H a_n+k_a a_w^n
\end{equation}
where $W_{\mathrm{H_2O}}$ represents the water uptake in grams H$_2$O per grams of dry membrane, a$_n$ is water activity, $A_L$, and $b_L$ are the Langmuir adsorption isotherm, $k_H$ is the Henry's Law constant, $k_a$ is the water aggregation reaction constant and $n$ is for water clustering. The model parameters 
are shown in Table \ref{tab:Park model parameters} along with the root mean squared error (RMSE) value to show the accuracy of the model's fit.
\begin{table} [hbt!]
    \centering
    \caption{Park model parameters for water concentration.}
    \label{tab:Park model parameters}
    \begin{adjustbox}{width=.7\textwidth}
    \begin{tabular}{|c|c|c|c|c|c|} \hline 
         $A_L$& $b_L$& $k_H$& $k_a$& $n$& $RMSE$\\ 
         \hline 
         0.02034& 92.0514& 0.1540& 0.2324& 5.8457&0.0068\\ 
         \hline 
    \end{tabular}
    \end{adjustbox}
\end{table}

Once $W_{\mathrm{H_2O}}$ is determined from Eq.~(\ref{Eq water concentration}), it can be converted to the concentration of water using:
\begin{equation}
\label{convert W to C}
C_{\mathrm{H_2O}} = \frac{W_{\mathrm{H_2O}} \times \left( \frac{10^6}{\mathrm{M}_{\mathrm{H_2O}}} \right)}{\frac{1}{\rho_{\text{dry membrane}}} + \frac{W_{\mathrm{H_2O}}}{\rho_{\mathrm{H_2O}}}}
\end{equation}
where $C_{\mathrm{H_2O}}$ is the water concentration in mol/m\textsuperscript{3}, $W_{\mathrm{H_2O}}$ is the mass of absorbed water, $\mathrm{M}_{\mathrm{H_2O}}$ is the molar mass of water, $\rho_{\text{dry membrane}}$ is the density of the dry membrane (g/cm\textsuperscript{3}), and $\rho_{\mathrm{H_2O}}$ is the density of water (g/cm\textsuperscript{3}).

\subsubsection{Hygroscopic Swelling Strain}   
\begin{figure}[htbp]
    \centering
    \begin{subfigure}[b]{0.6\textwidth}
        \centering
        \includegraphics[width=\linewidth]{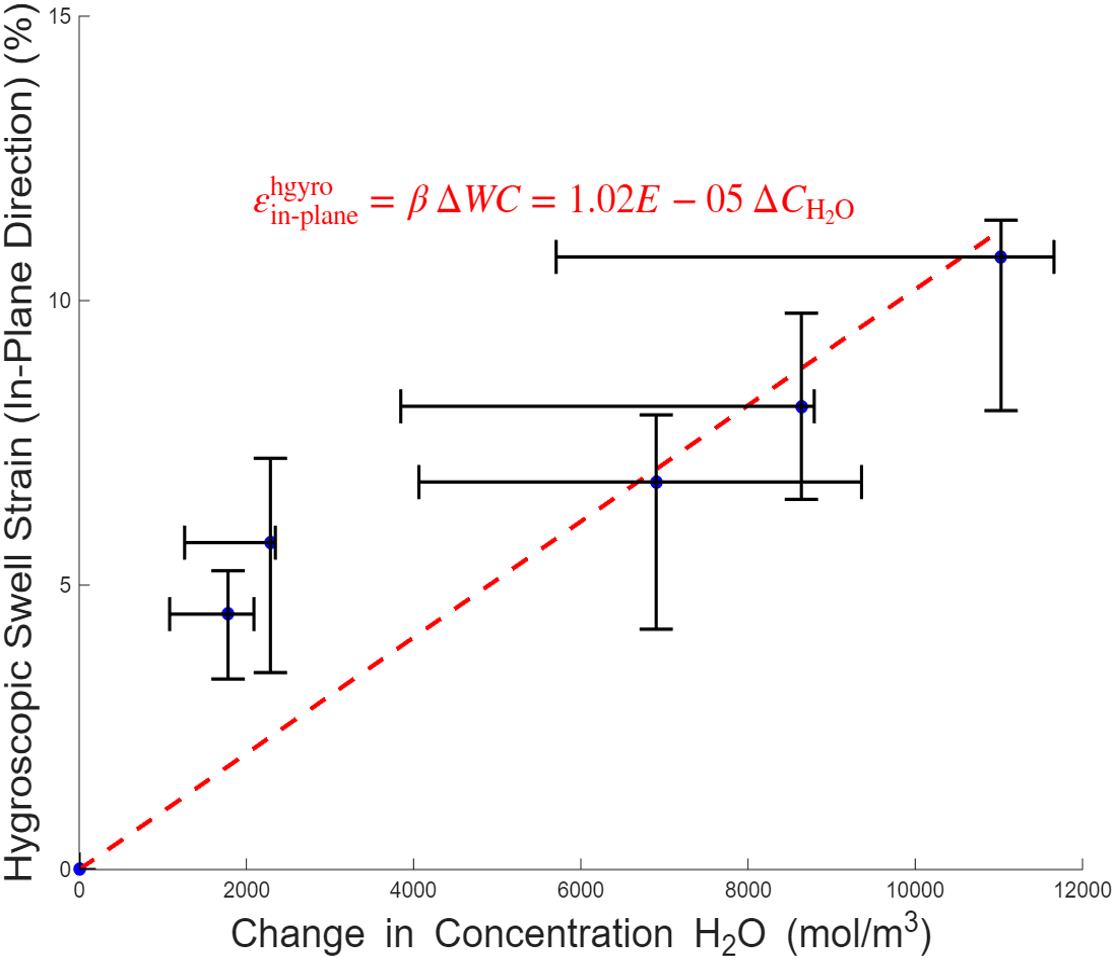}
        \caption{In-plane direction}
        \label{fig:swell_inplane}
    \end{subfigure}
    \hfill
    \begin{subfigure}[b]{0.6\textwidth}
        \centering
        \includegraphics[width=\linewidth]{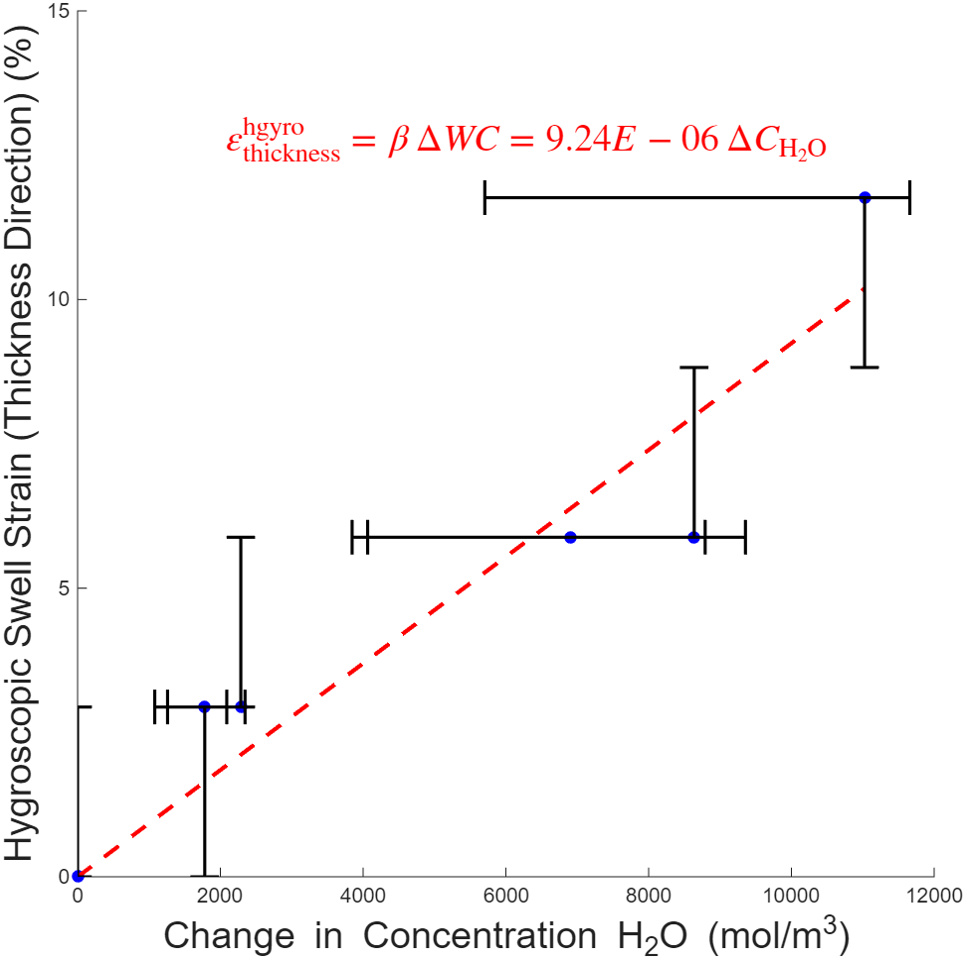}
        \caption{Thickness direction}
        \label{fig:swell_thickness}
    \end{subfigure}
    \caption{Hygroscopic swell strain versus the change in water concentration: (a) in-plane direction, (b) thickness direction from the equilibrated loading at 32\% RH.}
    \label{fig:swell_test_results}
\end{figure}

The strain of the sample at various humidities was determined in both the in-plane and thickness directions, at room temperature, \SIrange{18}{20}{\celsius}. The measured relative humidity was correlated with the water content using the data in Figure \ref{fig:con vs hum}. The resulting hygroscopic swell strain as a function of changes in water concentration levels is seen in Figure \ref{fig:swell_test_results}.
The data points shown in Figure \ref{fig:swell_test_results} represent the average measurement, and the solid black lines show the range of values measured for both strain and change in water content
The relationship between change in concentration and swell strain is approximately linear, and the swell coefficient ($\bm{\beta}$) was determined to be 1.02x10$^{-5}$ m$^3$/mol in the in-plane direction and 0.93x10$^{-5}$ m$^3$/mol in the thickness direction. 

Khalid et al. also measured the swell behavior of FAA-3, but they only captured two points (dry and wet), so a linear trend had to be used to fit the data \cite{Khalid}. The swell test conducted in this study confirmed this linear relationship. 

However, Khalid et al. found that the swelling strain was approximately 15\% and 43\% in the in-plane and thickness directions, respectively, at a temperature of \SI{30}{\celsius} \cite{Khalid}. We found a hygroscopic swell around 11\% in both directions at a change in concentration of 11026 mol/m$^3$ and a temperature of \SI{18}{\celsius}. 
This difference in the magnitude of swell could be related to the difference in temperature. 
As molecules vibrate more at higher temperatures, there is more room for water molecules to penetrate and swell the material, and thus, it should be expected that $\varepsilon^{\text{hygro}}$ increases as temperature increases.  This implies that $\beta$ should be a function of temperature.  However, this functional dependence was not measured in either this work or Khalid et al.'s work.

Potential reasons for the additional anisotropy observed by Khalid et al. may also be related to temperature, as well as measurement tools.  
FAA-3 may be more anisotropic at higher temperatures. However, it seems more likely that the material was softer with the increase in temperature, and thus, the thickness is more likely to be underestimated (and thus, strain is overestimated) due to the squeezing of the material while measuring. Future work may consider optical measurements of thickness swelling strain to avoid this potential measurement error.

\subsubsection{Excess Volume of Mixing}

Swelling measurements, combined with independently determined water solubility data, allow not only the determination of a swelling coefficient but also a deeper examination of the volume expansion of the membrane relative to the ideal volume of absorbed water. 
The excess volume of mixing captures the non-ideal volumetric interactions between sorbed water and the polymer matrix. In an ideal system, the total volume of the swollen membrane equals the sum of the dry polymer volume and the volume of the absorbed water, 18 cm$^3$/mol. In practice, however, swelling often leads to a total volume that is either smaller due to filling of the condensed water into non-equilibrium micro-voids of glassy polymers, or larger due to molecular rearrangements, i.e., polymer relaxation \cite{duan2013water,wu2011differences,raharjo2007pure}.

The membrane’s volume expansion upon swelling was calculated as:
\begin{equation}
\frac{\mathrm{d}V_{\text{swelling}}}{V_0} = (1 + \varepsilon^{\text{hygro}}_{\text{thickness}})(1 + \varepsilon^{\text{hygro}}_{\text{in-plane}})^2 - 1
\end{equation}
where $\varepsilon^{\text{hygro}}$ is the engineering strain due to swell. With knowledge of the water uptake, the excess volume can be determined from:
\begin{equation}
\frac{V_{\text{excess}}}{V_0} = \frac{\mathrm{d}V_{\text{swelling}}}{V_0} - \frac{\mathrm{d}V_{\text{H}_2\text{O}}}{V_0}
\end{equation}
and
\begin{equation}
\frac{\mathrm{d}V_{\text{H}_2\text{O}}}{V_0} = \frac{n_{\text{H}_2\text{O}} \bar{V}_{\text{H}_2\text{O}}}{V_0}
\end{equation}
where \(n_{\text{H}_2\text{O}}\) is the number of moles of absorbed water and \(\bar{V}_{\text{H}_2\text{O}}\) is the molar volume of pure water (taken as 18~cm\(^3\)/mol at room temperature). Positive values of \(V_{\text{excess}}\) indicate swelling beyond ideal mixing, while negative values suggest contraction relative to the sum of the components.\

\begin{figure}[hbt!]
        \centering
        \includegraphics[width=0.7\linewidth]{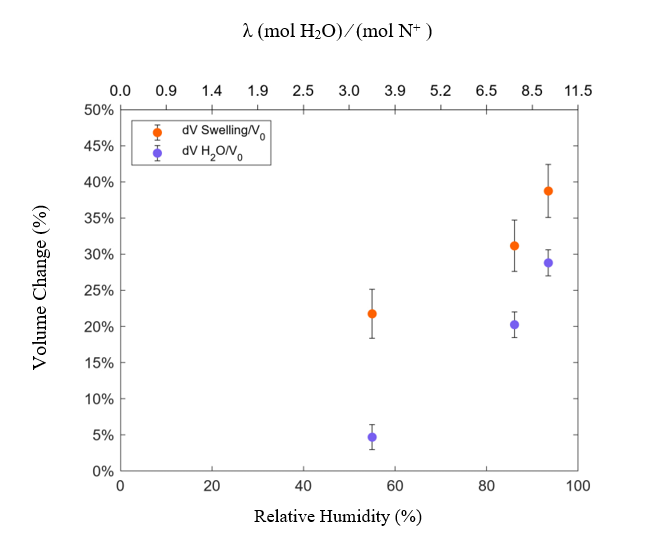}
        \caption{Measured volume change of Fumasep FAA-3 \(\left( \frac{\mathrm{d}V_{\text{swelling}}}{V_0} \right)\) as a function of relative humidity, compared with the expected volume change \(\left( \frac{\mathrm{d}V_{\text{H}_2\text{O}}}{V_0} \right)\) assuming ideal mixing, where each mole of water contributes 18~cm\(^3\).}
        \label{fig:volume_change}
    \end{figure}

Figure~\ref{fig:volume_change} shows the measured volume change of Fumasep FAA-3 as a function of water uptake, alongside the expected volume change assuming each mole of absorbed water occupies 18~cm\(^3\). 
Error bars represent the propagated uncertainty from both the swelling strain measurement and the standard deviation of triplicate water uptake measurements. These uncertainties were treated as independent and propagated through the calculations of \(\frac{\mathrm{d}V_{\text{swelling}}}{V_0}\) and \(\frac{\mathrm{d}V_{\text{H}_2\text{O}}}{V_0}\) using standard error propagation techniques.

Across the hydration range, the membrane exhibits greater volume expansion than expected from ideal mixing, indicating a positive excess volume of mixing. This implies that water uptake not only fills available free volume but also induces structural relaxation and polymer chain expansion, leading to an overall volumetric gain. The result suggests that the partial molar volume of water in the membrane exceeds that of bulk water under these conditions. While these findings point to strong non-ideal interactions between water and the polymer, it is important to note that air---including CO\textsubscript{2}---was always present, meaning that the results cannot be interpreted as a pure measurement of water’s partial molar volume. Gas uptake or displacement may also contribute to the observed expansion.

This finding is particularly noteworthy for an anion exchange membrane (AEM), where traditionally negative excess volumes of mixing have been reported. For example, A201, an AEM made by Tokuyama Corporation, displayed a contraction upon water sorption \cite{duan2013water}, and similarly, negative excess volumes have been measured for SPEEK-based cation exchange membranes~\cite{wu2011differences}. The swelling behavior of FAA-3 may have important implications for understanding hydration-driven changes in mechanical properties and selective transport performance in AEM applications.

\subsection{Thermal Expansion}
\label{sec:Temperature Results}
\begin{figure}[H]
    \centering
    \includegraphics[width=0.65\linewidth]{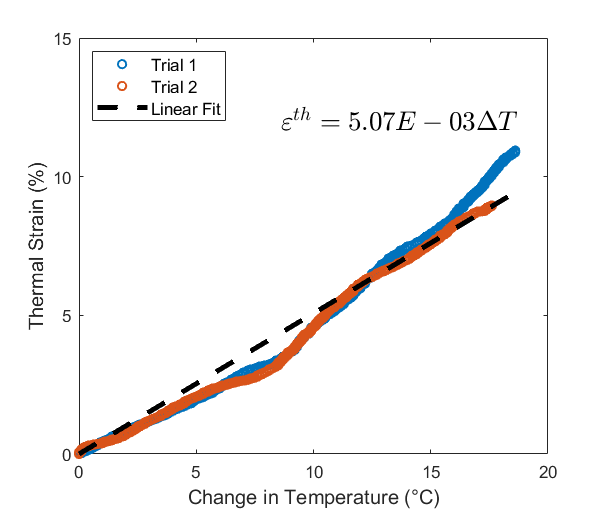}
    \caption{Thermal strain versus change in temperature with a linear fit.}
    \label{fig:thermal expansion}
\end{figure}

The relationship between the change in temperature and strain is shown in Figure \ref{fig:thermal expansion}, with the reference temperature being the ambient condition of \SI{21}{\celsius}.
When the temperature increases, the molecules vibrate more and move farther apart, resulting in thermal strain. The thermal expansion coefficient ($\alpha$) is determined through a linear fit of the experimental data (Figure \ref{fig:thermal expansion}). This was found to be approximately $5.07 \times 10^{-3}$/\SI{}{\celsius}. The standard error of fit (S$_{yx}$) for the linear curve was approximately 0.0055. In the work of Khalid et al., the material size increased with temperature, consistent with the trend observed in Figure \ref{fig:thermal expansion} \cite{Khalid}, however, the thermal expansion coefficient could not be found from their published data.It is important to note that this thermal expansion coefficient likely depends on relative humidity.  However, due to experimental limitations, this dependence was not explored.

\subsection{Ultimate Tensile Test}
\label{sec:Ultimate Tensile Results}
\subsubsection{Rate Variance}
\label{sec:Ultimate Tensile Rate Results}
\begin{figure}[hbt!]
        \centering
        \includegraphics[width=0.70\linewidth]{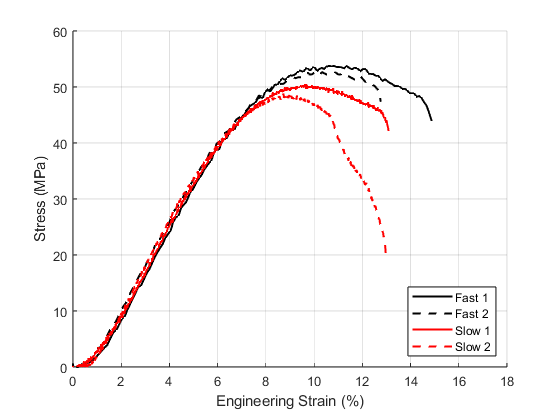}
        \caption{Ultimate tensile test nominal stress versus the mechanical part of the engineering strain at engineering strain rates of 0.1 s$^{-1}$ (fast) and 0.01 s$^{-1}$ (slow), at \SI{23}{\celsius} and 42\%RH.}
        \label{fig:Rate Ultimate Tensile Test eng}
    \end{figure}

The uniaxial tensile behavior of FAA-3 at ambient conditions (approximately \SI{23}{\celsius}, 41-43\%RH) is shown in Figure \ref{fig:Rate Ultimate Tensile Test eng}. The stress-strain curve contains a linear region at strains less than approximately 6\%, followed by a nonlinear region which includes the maximum stress experienced by the material at around 50--55 MPa, and fracture occurs around 13--14\% strain.    

The strain rate at which the sample is loaded influences the maximum stress, with faster strain rates leading to slightly higher stress. This is expected since viscous effects lead to stress relaxation with time. Despite the change in strain rate, the initial linear region and the general shape of the graphs were similar. Table \ref{tab:Ultimate Tensile Test} provides the values of key parameters such as maximum stress, the slope of the linear region (Young’s modulus), and elongation at break under various conditions. 

\begin{table} [hbt!]
    \centering
    \caption{Mechanical properties of FAA-3 found from the ultimate tensile test.}
    \label{tab:Ultimate Tensile Test}
    \begin{adjustbox}{width=\textwidth}
    \begin{tabular}{|c|c|c|c|c|c|c|c|} \hline 
         & Fast 1& Fast 2& Slow 1& Slow 2 & Humidified 1& Humidified 2& Humidified 3\\ 
         \hline 
         Rate (s$^{-1}$):& 0.1& 0.1& 0.01& 0.01& 0.01& 0.01& 0.01\\ 
         \hline 
         Humidity (\%):& 43.2& 42.2& 41.2& 41.4& 67.9& 71.8& 73.5\\ 
         \hline 
         Temperature (\SI{}{\celsius}):& 22.9& 23.2& 23.4& 23.5& 22.2& 22.1& 22.0\\ 
         \hline 
                  Maximum stress (MPa):& 53.8& 52.7& 50.5& 48.8& 41.3& 34.5& 32.3\\ 
         \hline 
              Young’s Modulus (MPa):& 694& 693& 696& 704& 486& 404& 383\\ 
         \hline 
         Mechanical strain at break (\%):& 14.9& 12.8& 13.1& 13.0& 21.0& 12.9& 13.5\\ 
         \hline
         Swell (\%):& N/A& N/A& N/A& N/A& 7.03& 6.7& 8.22 \\ 
         \hline
    \end{tabular}
    \end{adjustbox}
\end{table}

The Young’s Modulus was close to 700 MPa regardless of the strain rate. 
The ultimate tensile strength was around 53 MPa for the faster strain rate of 0.1 s$^{-1}$ and around 50 MPa for the slower strain rate of 0.01 s$^{-1}$. The average strain the sample reached before fracturing was 13.82\% at the faster rate and 13.06\% at the slower rate. These results show that in the range of rates tested, the strain rate had little influence on the overall tensile properties of FAA-3 since the values were relatively similar for all the properties. 


\subsubsection{Humidity Variance}
\label{sec:Ultimate Tensile Humidity Results}
\begin{figure}[hbt!]
        \centering
        \includegraphics[width=0.7\linewidth]{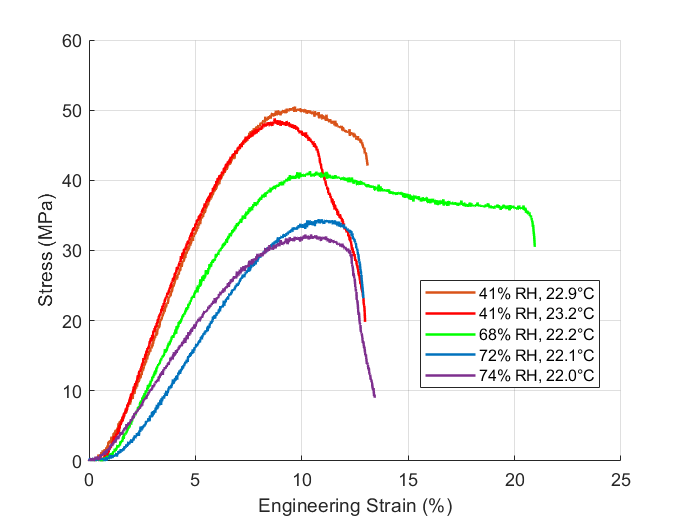}
        \caption{Ultimate tensile test results for nominal stress versus mechanical part of the engineering strain at various relative humidity levels and room temperature. }
        \label{fig:Humidity Ultimate Tensile Test eng}
    \end{figure}

Although the mechanical behavior of FAA-3 did not vary much with strain rate, the change in relative humidity has a noticeable impact on the uniaxial stress-strain response, as seen in Figure \ref{fig:Humidity Ultimate Tensile Test eng}.

Although both the ambient and high relative humidity samples still contain a linear region and a non-linear region, the maximum stress is decreased significantly, and the linear region has a smaller slope when humidity is increased. This is expected, as higher humidity alters the mechanical properties of many materials, such as wood~\cite{wood} and polymers like poly(methyl methacrylate) (PMMA)~\cite{PMMA}. When water is absorbed into the material, it acts as a plasticizer, lubricating the polymer chains and disrupting the bonds between molecules, which can reduce both stiffness and strength~\cite{betts2013anatomy,clark2018biology}. 

Table~\ref{tab:Ultimate Tensile Test} provides the values of each key parameter obtained from Figure~\ref{fig:Humidity Ultimate Tensile Test eng}. Increasing humidity caused a decrease in maximum stress and Young’s Modulus, while the mechanical strain at break was generally similar for all humidity levels. This consistent engineering strain at break suggests that the failure of the material might stem from breaking the cross-link bonds. These cross-link bonds act as fixed points within the polymer chain and are the most likely sites to fail when the material is stretched to its limit, irrespective of the surrounding moisture conditions. Note that the first high humidity test did show significantly more elongation at break, but additional testing was unable to confirm this, and there is insufficient data to determine if this was an outlier. Also note that the elongation at break in Table~\ref{tab:Ultimate Tensile Test} is the mechanical engineering strain at break. The overall strain the sample experiences at high RH is much greater than under ambient conditions, since the sample elongated by as much as 7\% before any mechanical load was applied due to hygroscopic swelling.

The trends in the results correlate with the FAA-3 testing by Khalid et al., who found decreasing tensile strength, decreasing Young's modulus, and increasing elongation at break with increasing humidity and temperature for FAA-3 \cite{Khalid}. Along with this behavior, Khalid et al. found that FAA-3 will become more ductile, making it more likely to stretch to a longer length before
breaking as temperature and humidity increase. Although the trends among the results are generally consistent, the values obtained for each of the properties differ significantly. Table \ref{tab:Property comparison} compares our results with other studies on various anion exchanged membranes.

\renewcommand{\arraystretch}{1.5}
\begin{table}[H]
    \centering
    \caption{Mechanical properties of anion exchange materials.}
    \label{tab:Property comparison}
    \begin{adjustbox}{width=\textwidth}
    {
        \Large 
        \begin{tabular}{|c|c|c|c|c|c|c|c|}
        \hline 
        Author & Material & Strain  Rate (s$^{-1}$): & Humidity (\%RH): & Temperature (\SI{}{\celsius}): & Maximum stress (MPa): & Young’s Modulus (MPa): & Max Mechanical Strain (\%) \\
        \hline
        This Work & FAA-3-30-HCO$_3$ & 0.01 & 42 & 23 & 53 & 700 & 13 \\
        \hline
        This Work & FAA-3-30-HCO$_3$ & 0.01 & 72 & 23 & 33 & 400 & 13 \\ 
        \hline 
        Khalid et al. \cite{Khalid} & FAA-3-50-Cl & 0.004 & 50-70 & 23-27 & 16 & 520 & 40 \\ 
        \hline 
        Khalid et al. \cite{Khalid} & FAA-3-50-Cl & 0.004 & 100 & 30 & 14 & 75 & 100 \\ 
        \hline 
        Kushner et al. \cite{kushner2016side} & QA-PPO: BTMA40 & - & 40 & 24 & - & 1000 & - \\ 
        \hline 
        Kushner et al. \cite{kushner2016side} & QA-PPO: BTMA40 & - & 70 & 24 & - & 700 & - \\ 
        \hline
        Zheng et al. \cite{ZHENG2024854} & c(Je)-QAPVB-6.25 AEM & - & 70 & 25 & 13 & - & 15 \\
        \hline
        Zheng et al. \cite{ZHENG2024854} & c(Je)-QAPVB-12.5 AEM & - & 70 & 25 & 6 & - & 18 \\
        \hline
        Narducci et al. \cite{narducci2016mechanical} & PSU-TMA-HCO$_3$ & - & 50 & 25 & 44 & 985 & 10 \\
        \hline
        Narducci et al. \cite{narducci2016mechanical} & PSU-TMA-OH & - & 50 & 25 & 20 & 680 & 11 \\
        \hline
        \end{tabular}
    }
    \end{adjustbox}
\end{table}

While the values between each anion exchange material differ from one another, Table \ref{tab:Property comparison} shows that anion exchange membranes generally have a Young's modulus between 500-1000~MPa at \SI{25}{\celsius} and 40\%RH. For FAA-3, the tensile strength in dry conditions (50-70\%RH and \SIrange{23}{27}{\celsius}) from Khalid et al. was much smaller than the experimental results in this work at 72\%RH and \SI{23}{\celsius}, while the elongation at break in dry conditions from Khalid et al. was much greater than the experimental results in this work. The Young's modulus at 70\%RH were similar. The other values Khalid et al. found could not be used for comparison as their experiments were done while changing both the water content and temperature simultaneously. The load rate applied by Khalid et al. was at a rate of 0.004~s$^{-1}$ (0.17~mm~s$^{-1}$), which is slower than the 0.01~s$^{-1}$ (0.12~mm~s$^{-1}$) applied in our experiment, but likely not enough to explain the difference is strength and elongation. Note that the strain rate was not specified by Khalid et al., thus the strain rate of 0.004 s$^{-1}$ in Table \ref{tab:Property comparison} was calculated using the dimensions reported by Khalid et al. (4 $\times$ 1~cm$^2$) and assuming that 4 cm was the length in the direction of load.

The discrepancies in material properties became even more pronounced when compared to the other materials listed in Table \ref{tab:Property comparison}, as those experiments do not utilize FAA-3. Nonetheless, these results show that the Young's modulus, strength, and elongation at break are similar for FAA-3 compared to other anion exchange materials. FAA-3 does have a higher strength than other AEMs, but that may be because the tests were performed in Arizona, USA, where the ambient relative humidity is much lower than most temperate climates, and indeed those who reported lower strength all did so at higher relative humidities. It was also shown by Narducci et al. \cite{narducci2016mechanical} that bicarbonate exchanged materials have higher strength and stiffness values compared to other forms of anion exchanged materials, which could support the high values seen in bicarbonate exchanged FAA-3. The maximum mechanical strain of FAA-3 differed from Khalid et al. but was in line with others. While differences in elongation for the submerged tests can be explained by the water disrupting the bonds and lubricating slip between the chains, Khalid et al. also saw significantly more elongation than others for dry materials.

\subsection{Load-Hold-Unload Test}
\label{sec:Load-Hold-Unload Results}
\subsubsection{Humidity Variance}
\label{sec:Load-Hold-Unload Humidity Results}
\begin{figure}[htbp]
    \centering

    \begin{subfigure}[b]{0.7\linewidth}
        \centering
        \includegraphics[width=\linewidth]{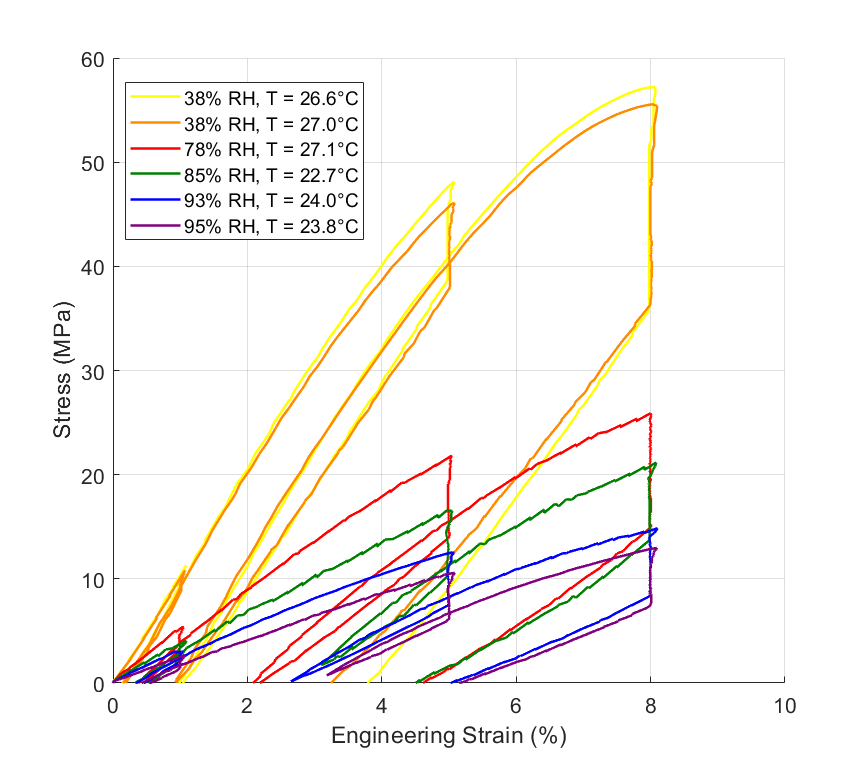}
        \caption*{(a) Stress vs.\ engineering strain}
    \end{subfigure}

    \vskip\baselineskip

    \begin{subfigure}[b]{0.7\linewidth}
        \centering
        \includegraphics[width=\linewidth]{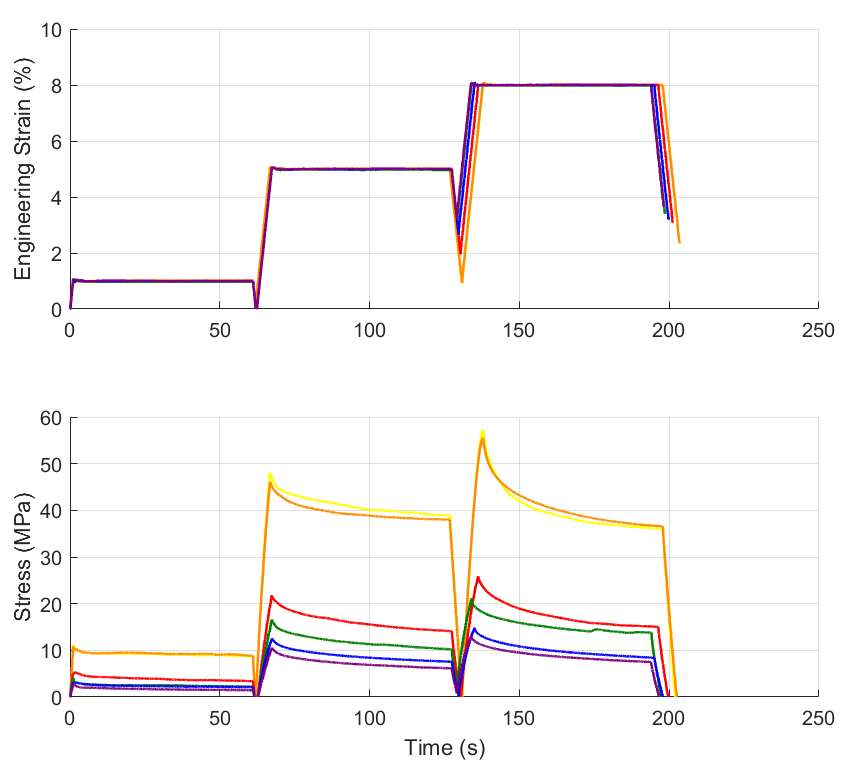}
        \caption*{(b) Engineering strain vs.\ time (top) and stress vs.\ time (bottom)}
    \end{subfigure}
    \caption{Mechanical behavior of the membrane under varied relative humidity conditions during load-hold-unload testing.      (a) Stress vs. engineering strain curves compare different RH levels at room temperature. (b) Engineering strain vs.\ time (top) and nominal stress vs.\ time (bottom) illustrate viscoelastic relaxation.}
    \label{fig:Humidity_LHU}
\end{figure}

The results from the load-hold-unload test in Figure \ref{fig:Humidity_LHU} show the effects the relative humidity has on mechanical behavior, including viscous effects (stress relaxation during strain holds) and plastic deformation (residual strains after unloading). Note that because the unload was not held, the plastic deformation seen may be partially or fully recovered over time.  

The experimental results were at ambient temperature (ranging from \SIrange{23}{27}{\celsius}) and at a low, medium, and high relative humidities. However, due to a lack of precise control of humidity, there is a range for each category: low or ambient RH is 35-40\%RH, medium RH is 75-85\%RH, and high RH is 90-95\%RH. The load of the load-hold-unload test reveals the same trend as the ultimate tensile test. 
The Young’s Modulus is around 1007-1011 MPa in ambient RH and high RH (90\%-95\%) is around 244-332 MPa. These values are all within range of others (see Table \ref{tab:Property comparison}), and trends seen previously (section \ref{sec:Ultimate Tensile Results}). Although the ambient modulus increased compared to the ultimate tensile test, this may be due to the lower ambient humidity of 37\%RH in these load-hold-unload tests compared to the ambient humidity of 42\%RH in the ultimate tensile test. This emphasizes the importance of relative humidity on the mechanics of the material, since small changes in RH seem to cause measurable variations in the mechanical properties.

The hold component examines the stress relaxation that occurs when held at constant strain. 
The initial stress relaxation is the greatest, and the relaxation gradually levels out with time.  Within one minute, the relaxation is nearly complete for all strain levels tested.
The relaxation is most notable at higher strains. This is because the stress is higher, so there is more to relax. 
The amount of stress that decreased was much greater at ambient RH than at high RH. 
Again, this is simply due to larger stresses in ambient conditions leading to more stress that can be relaxed.  

The unload assesses any potential residual (plastic) deformation. When the loaded sample was released to a force close to zero (but non-zero to keep the sample taut), it was found that the plastic deformation of FAA-3 increased with increasing RH. The sample had little residual deformation in ambient conditions and was almost able to return to its original length after being strained to 1\%. With an increase in mechanical strain to 5\% and 8\%, the amount of residual deformation increased, and the sample did not return to its original length. These trends remained as the relative humidity was increased, but the plastic deformation was much greater. As seen in Figure \ref{fig:Humidity_LHU}, the residual strain in the sample when the force returned to a very small value was larger at higher relative humidity values. Again, this may be due to water lubricating the chains and disrupting some bonds, making slipping between chains easier.

These results are significant since FAA-3 will be loaded axially in DAC for continuous separation by clipping the edges of the membrane and making one side of the membrane wet and the other dry. This will result in bending as the wet side swells while the dry side does not. At the site where the FAA-3 is held in place, the material will want to swell, but cannot. This will result in significant stresses at these locations, but the stresses should relax with time. Then, when the DAC process is stopped or paused, there will be residual deformation of the material. These results show how much plastic deformation may differ depending on the water concentration and the time scale for relaxation.

\subsubsection{Temperature Variance}
\label{sec:Load-Hold-Unload Temperature Results}

\begin{figure}[htbp]
    \centering

    \begin{subfigure}[b]{0.7\linewidth}
        \centering
        \includegraphics[width=\linewidth]{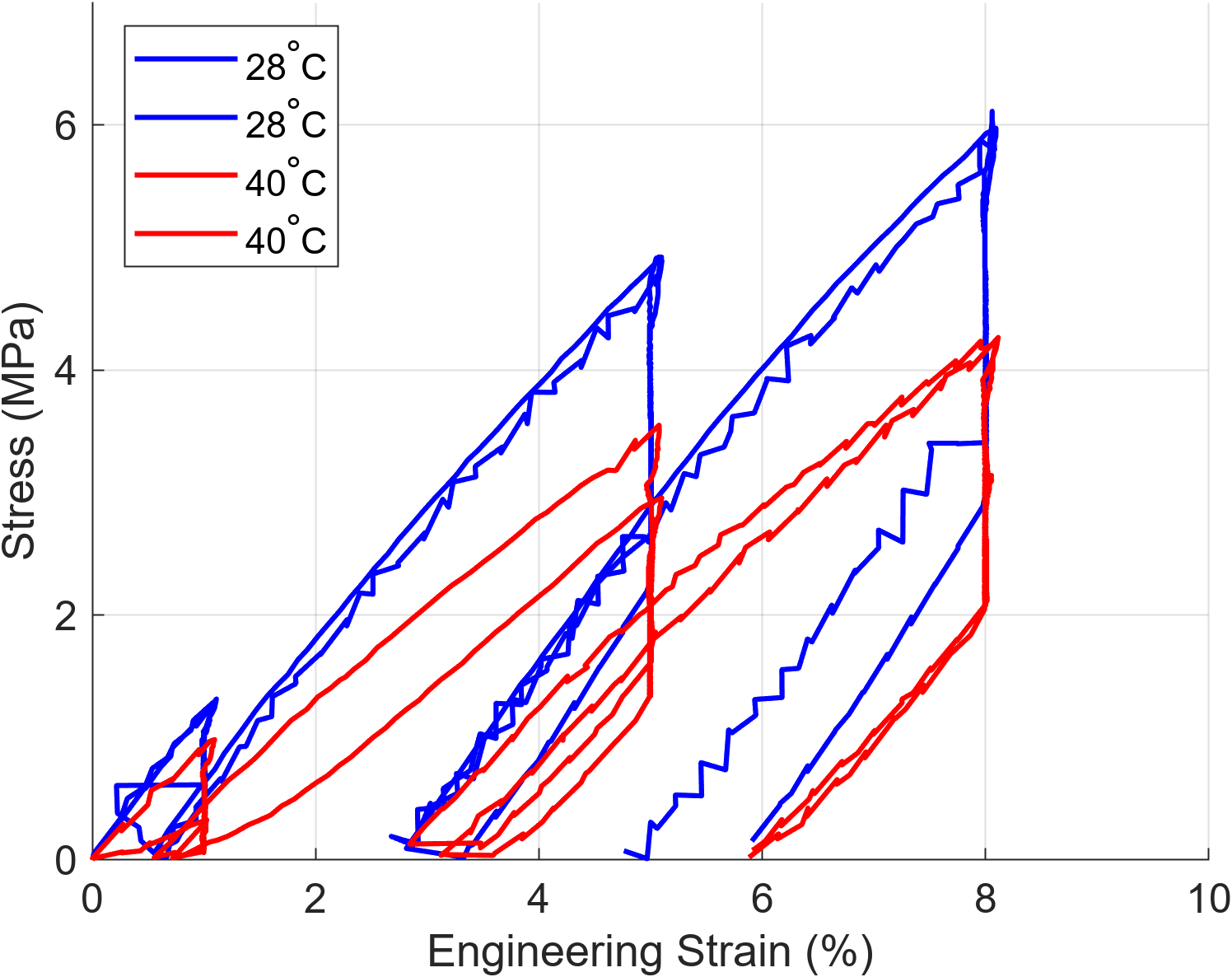}
        \caption*{(a) True stress vs.\ true strain}
    \end{subfigure}

    \vskip\baselineskip

    \begin{subfigure}[b]{0.7\linewidth}
        \centering
        \includegraphics[width=\linewidth]{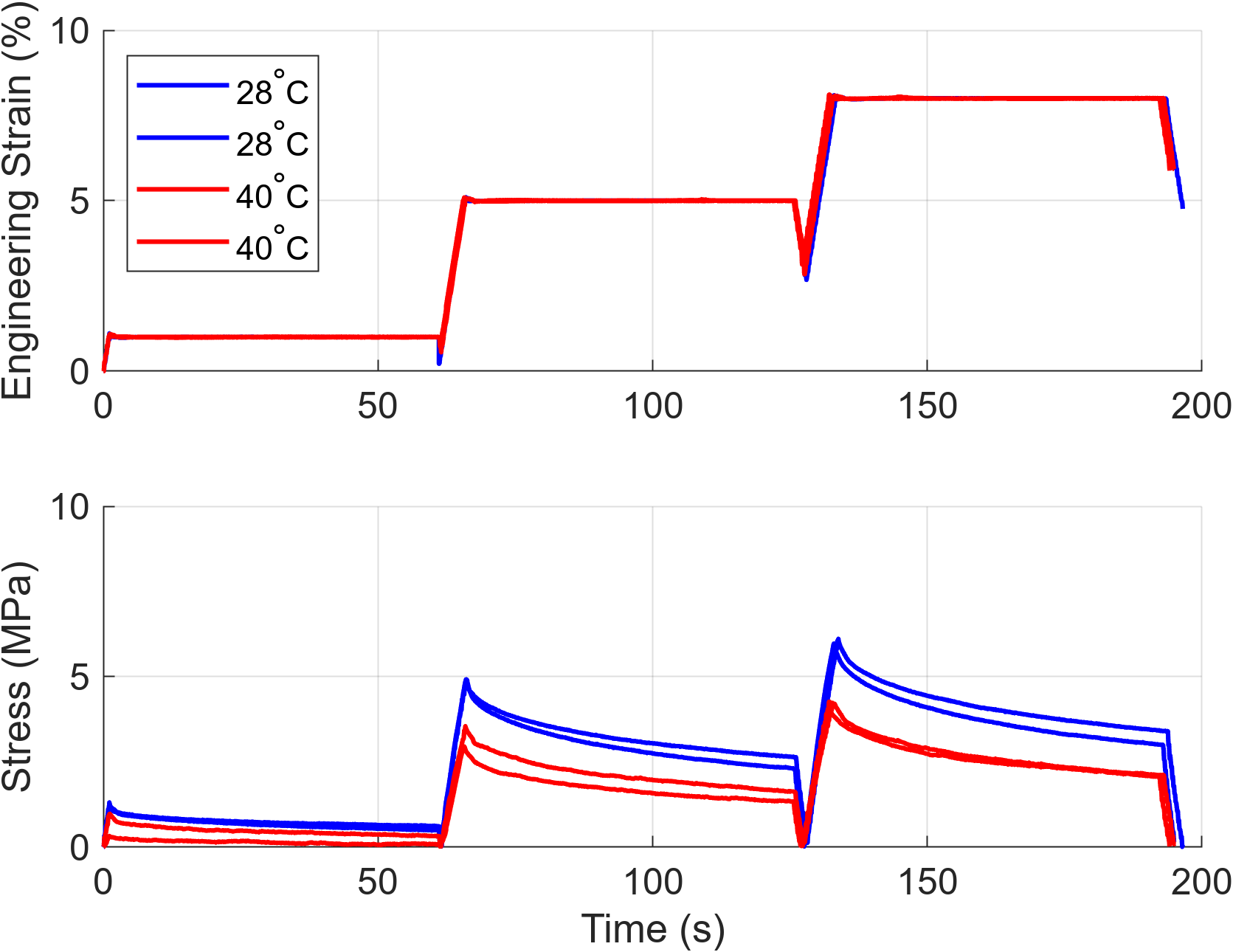}
        \caption*{(b) True strain vs.\ time (top) and true stress vs.\ time (bottom)}
    \end{subfigure}

    \caption{Mechanical response of the membrane at two temperatures during load-hold-unload testing: (a) Nominal stress vs.\ engineering strain curves show the deformation behavior at 28°C and 40°C; (b) Engineering strain vs.\ time (top) and nominal stress vs.\ time (bottom) illustrate viscoelastic relaxation. Both trials are included for each temperature.}
    \label{fig:Temp_LHU}
\end{figure}

The influence of temperature was examined through the load-hold-unload tests of FAA-3 submerged in DI water. By submerging the sample, the relative humidity is kept constant at 100\%RH, and the temperature effects on the mechanical behavior of FAA-3 can be isolated and measured.  The results of these tests are shown in Figures  \ref{fig:Temp_LHU}.

Like the results of the load-hold-unload test in the humidity variance (Section \ref{sec:Load-Hold-Unload Humidity Results}), the load reveals the Young's modulus, the hold shows stress relaxation, and the unload shows plastic deformation, although some or all of the plastic deformation seen may be recovered over time. The trends shown with increasing humidity were also prevalent with increasing temperature. As seen in Figure \ref{fig:Temp LHU}, the stiffness decreases when the temperature is increased. 
This is consistent with the DMA results since the storage modulus decreased with the rising temperature.

The hold component shows a similar trend as seen with humidity, more stress relaxation at higher strains.   
The change in stress relaxation was relatively small, but a decrease was observed when the temperature was increased. However, when comparing the change in stress relaxation between varying temperatures and varying humidity, it can be seen that the change in humidity has a much greater impact. 
This shows that for the conditions in which FAA-3 will likely be used, humidity has a much larger influence on stress relaxation than temperature. This implies that material parameters associated with stress relaxation and viscous effects may not need to be temperature-dependent. 

The unload assesses any residual (plastic) deformation in the polymer membrane. When the loaded sample was unloaded to a force near zero, it was found that plastic deformation increased with increasing temperature as expected, since the slipping of molecules is easier at higher temperatures, but the influence of temperature on the plastic deformation was small. Figure \ref{fig:Temp_LHU} shows significant plastic deformation, but the plastic deformation does not differ much between the varying temperatures. This may be because the range of temperatures is relatively small.  

  \begin{figure}[htbp]
    \centering

    \begin{subfigure}[b]{0.7\linewidth}
        \centering
         
        \includegraphics[width=\linewidth]{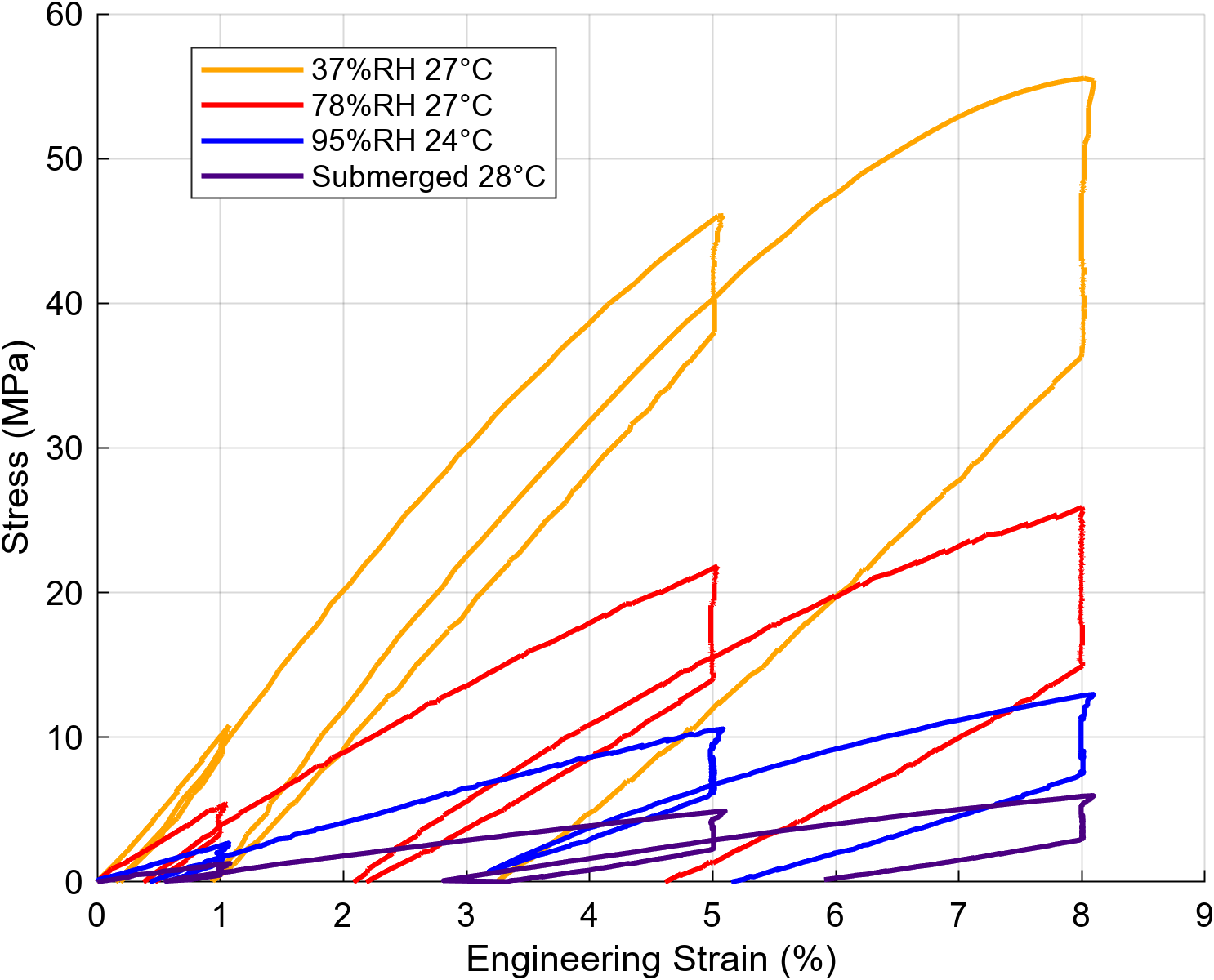}
        \caption*{(a) Nominal stress vs.\ engineering strain}
    \end{subfigure}

    \vskip\baselineskip

    \begin{subfigure}[b]{0.7\linewidth}
        \centering
        \includegraphics[width=\linewidth]{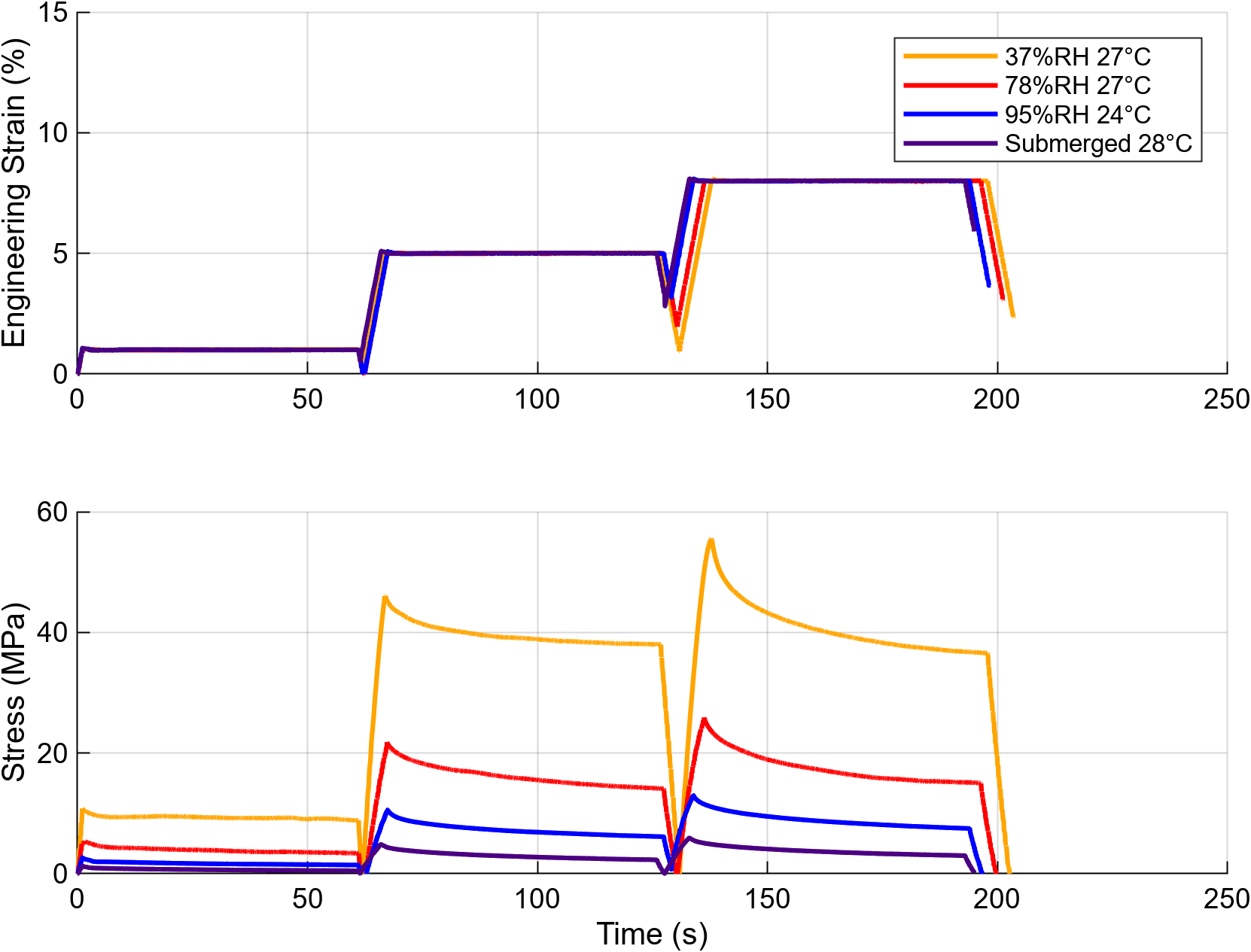}
        \caption*{(b) Engineering strain vs.\ time (top) and nominal stress vs.\ time (bottom)}
          
    \end{subfigure}

    \caption{Load-hold-unload testing of membrane under different humidity conditions: (a) Nominal stress vs.\ engineering strain highlights the mechanical response during loading and unloading; (b) Engineering strain vs.\ time (top) and nominal stress vs.\ time (bottom) illustrate viscoelastic relaxation.}
    \label{fig:Combined_LHU}
\end{figure}

Figure \ref{fig:Combined_LHU}, shows how the submerged results correlate with the previous trends observed in the humidity variance results at similar temperatures. As the relative humidity increased until the sample was submerged, the stiffness of FAA-3 decreased. While the trends are as expected, the effects of submersion are significant. The Young's modulus of the material at 95\%RH was 244.86 MPa, and this decreased to approximately 41.85 MPa when submerged. 
This suggests that submerging a sample is not the same as a 100\% RH environment in terms of mechanical behavior. A possible explanation for this is Schroeder's paradox, that polymers have different maximum water uptake in the liquid and saturated vapor phases \cite{CHEN2023113050}, and this difference in water uptake would affect how the water molecules change the polymer bonds, which would, in turn, change the mechanical behavior.

\subsection{Cyclic Loading Test}
\label{sec:Cyclic Loading Results}
\begin{figure}[H]
        \centering
        \includegraphics[width=0.7\linewidth]{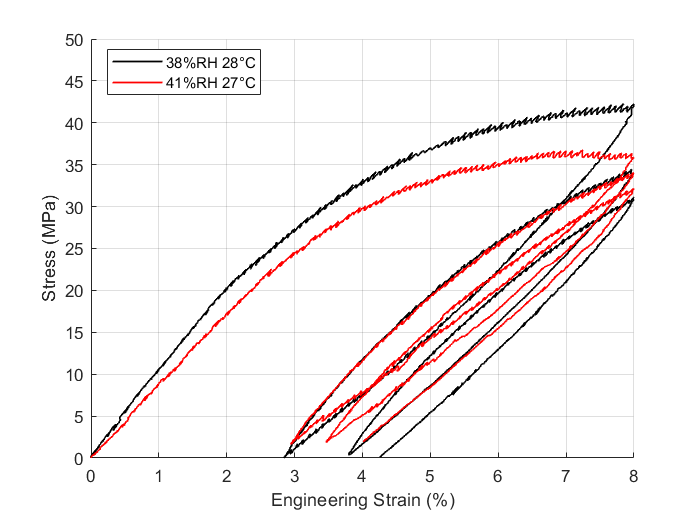}
        \caption{Cyclic loading results at \SI{27}{\celsius}, 38--41\%RH, and a strain rate of 0.0001 s$^{-1}$.}
        \label{fig:Cyclic Loading Test 1 eng}
    \end{figure}

Figure \ref{fig:Cyclic Loading Test 1 eng} shows the results from the cyclic loading tests, where the sample was loaded to 8\% engineering strain at an engineering strain rate of 0.0001 s$^{-1}$ and then unloaded to a force near zero, for a total of 3 cycles. 

As seen in the loading of the first cycle of Figure \ref{fig:Cyclic Loading Test 1 eng}, the stress increases with strain, but while the strain is increased at a constant rate, the stress is not linear near the end of the loading, likely because of viscous effects and/or plastic deformation. Stress relaxation may be present due to the slow strain rate.  Thus, the stress to both increases due to the applied force and decreases due to relaxation, and therefore, the stress begins to level out. When unloading the sample to a stress of approximately zero, the stress and strain are approximately linear, as is typical of elastic unloading. Although as the stress reaches nearly 0 MPa, the strain does not return to 0, indicating plastic deformation. The developed model should be able to simulate these viscous effects and plastic strains, and these cyclic loading results will be used to verify the accuracy of the model.

A comparison of the first and second trials reveals differences in material behavior, which can be attributed to variations in relative humidity and temperature conditions. The first trial was conducted at a lower relative humidity and higher temperature compared to the second due to different ambient conditions on the days the tests were performed. This difference was evident in the resulting curves, which showed the material in the first trial to be stiffer.  Higher stiffness is associated with lower temperature and RH, but this test was at a higher temperature and lower RH.  This suggests that the influence of RH on the elastic properties of FAA-3 is stronger than that of temperature. Additionally, there were slight variations in plastic deformation between the two trials. The first trial exhibited more plastic deformation than the second, suggesting that temperature has a more pronounced effect on plastic deformation than RH. Higher RH tends to increase plastic deformation, while lower temperatures reduce it. Since the second trial, conducted at a lower temperature, exhibited less plastic deformation despite a higher RH, it can be inferred that temperature exerts a stronger influence on the plastic deformation of the material.

\section{Material Modeling }
\label{sec:model}

\begin{figure}[H]
    \centering
    \includegraphics[width=0.7\textwidth]{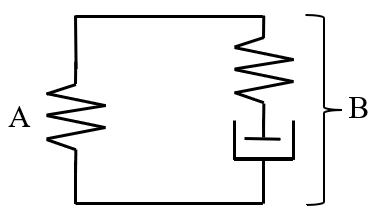} 
    \caption{Rheological representation of the Parallel network model, with Network A representing the elastic (Neo-Hookean spring) and Network B representing the viscous effects through an elastic element (Neo-Hookean spring) in parallel with a  Bergstrom-Boyce dashpot}
    \label{fig:BB_Model}
\end{figure}

The experimental data highlighted in the previous section reveal the complex, nonlinear, and time-dependent mechanical behavior of FAA-3, influenced significantly by relative humidity and temperature. To effectively capture these characteristics in a mathematical model, this study adopts a parallel-network framework inspired by the Bergstr\"{o}m-Boyce (BB) model \cite{bergstrom1998constitutive}.  The BB model framework was chosen because it has been used to successfully model AEMs, including their temperature and hydration dependence \cite{YOON20113933}. 
The model comprises two networks: Network A, which consists solely of an elastic spring, and Network B, which features an elastic spring in series with a viscous dashpot, as depicted in Figure \ref{fig:BB_Model}.
Network A’s spring represents the instantaneous elastic response due to polymer chain stretching, while Network B’s spring-dashpot combination accounts for both the immediate elastic deformation and the time-dependent viscous flow. 
Rather than using the full BB model, we found that reasonable results could be obtained with a simpler model, where both springs are considered Neo-Hookean and the dashpot is a BB flow element.  

In this parallel-network configuration, the total deformation gradient \( \mathbf{F} \) is uniformly applied across both networks, such that \( \mathbf{F} = \mathbf{F}_A = \mathbf{F}_B \). For Network B, the deformation gradient is decomposed into elastic and viscous components:
\begin{equation}
\mathbf{F}_B = \mathbf{F}_B^e \mathbf{F}_B^v.
\label{Eq F}
\end{equation}
Since the networks operate in parallel, the total Cauchy stress is the sum of the contributions from each network:
\begin{equation}
\bm{\sigma} = \bm{\sigma}_A + \bm{\sigma}_B.
\end{equation}
\subsection{Neo-Hookean Elastic Spring}

\label{subsec:neo_hookean}

The elastic behavior of the springs in both Network A and Network B is modeled using the Neo-Hookean hyperelastic formulation, which is well-suited for describing the nonlinear elasticity of polymer networks \cite{bergstrom1998constitutive}. 
The material is assumed incompressible (bulk modulus is infinite) following \cite{SILBERSTEIN20105692,YOON20113933,bergstrom1998constitutive}, as is typical for polymer membranes.  
The strain energy density function for an incompressible Neo-Hookean material is expressed as:
\begin{equation}
W = \frac{\mu}{2} (\bar{I}_1 - 3),
\end{equation}
where \( \mu \) is the shear modulus (\( \mu_1 \) for Network A and \( \mu_2 \) for Network B), and \( \bar{I}_1 \) is the first invariant of the deviatoric left Cauchy-Green deformation tensor, defined as:
\begin{equation}
\bar{I}_1 = \text{tr}(\bar{\mathbf{B}}),
\end{equation}
with \( \bar{\mathbf{B}} = J^{-2/3} \mathbf{F} \mathbf{F}^\top \), and \( J = \det(\mathbf{F}) \).  In this case, the material is assumed incompressible, so $J = 1$. The Cauchy stress tensor is then given by:
\begin{equation}
\boldsymbol{\sigma} = -p \mathbf{I} + \mu \mathbf{B},
\label{Eq neo hookean}
\end{equation}
where \( p \) is the hydrostatic pressure (a Lagrange multiplier for incompressibility), \( \mathbf{I} \) is the identity tensor, and \( \mathbf{B} = \mathbf{F} \mathbf{F}^\top \) is the left Cauchy-Green deformation tensor. For Network A, \( \mathbf{B} = \mathbf{F} \mathbf{F}^\top \) because \( \mathbf{F}_A = \mathbf{F} \).  For Network B, \( \mathbf{B} = \mathbf{F}_B^e {\mathbf{F}_B^e}^\top\) because only the elastic part of the deformation response is used in the Neo-Hookean spring. 

\subsection{Bergstr\"{o}m-Boyce (BB) Viscous Dashpot}
\label{subsec:bb_model}

The viscous response in Network B is modeled using a modified version of the Bergstr\"{o}m-Boyce (BB) flow element, which excels at capturing the time-dependent deformation of polymers \cite{bergstrom1998constitutive}. 
In this model, the viscous part of the rate of deformation is defined through a constitutive assumption that 
\begin{equation}
\mathbf{D}_B^v = \dot{\gamma}_B^v \mathbf{N}_B, 
\label{Eq Dv}
\end{equation}
where \( \dot{\gamma}_B^v \) is the viscous shear rate which defines the magnitude of $\mathbf{D}_B^v$, and 
\begin{equation}
\mathbf{N}_B = \frac{\text{dev}[\boldsymbol{\sigma}_B]}{\tau}
\label{Eq N_B}
\end{equation}
specifies the flow direction.  In Eq. (\ref{Eq N_B}), $\tau$
is the magnitude of the effective deviatoric stress in Network B, given by:
\begin{equation}
\tau = \|\text{dev}[\boldsymbol{\sigma}_B]\|_F              =\sqrt{tr[\boldsymbol{\sigma_B'}\boldsymbol{\sigma_B'}]}
\label{Eq tau}
\end{equation}
where $\boldsymbol{\sigma_B'}$ is the deviatoric part of $\boldsymbol{\sigma_B}$.  
Note that while Eq. (\ref{Eq Dv}) is general and 3-dimensional, all experimental tests were 1-dimensional, and therefore the direction of viscous flow expressed by Eq. (\ref{Eq N_B}) can not be experimentally verified in this work.  

Eq. (\ref{Eq Dv}) can be used to find the viscous part of the rate of the deformation gradient as:
\begin{equation}
\dot{\mathbf{F}}_B^v = \dot{\gamma}_B^v (\mathbf{F}_B^e)^{-1} \mathbf{N}_B \mathbf{F}_B^e \mathbf{F}_B^v.
\end{equation}
The exact form of viscous shear rate, $\dot{\gamma}_B^v$, needs to be defined to update the viscous deformation gradient, $\mathbf{F}_B^v$. 

The version of the  Bergstr\"{o}m-Boyce (BB) flow model used in this work uses the following viscous shear rate:
\begin{equation}
\dot{\gamma}_B^v = \dot{\gamma}_0 \left( \lambda_B^v - 1 + \xi \right)^c \left( \frac{\tau}{\hat{\tau}} \right)^m,
\end{equation}
where:
\begin{itemize}
    \item \( \dot{\gamma}_0 = 1 \, \text{s}^{-1} \) is a reference shear rate for dimensional consistency,
    \item \( \lambda_B^v \) is the viscous stretch ratio (Mises equivalent stretch), given by 
    \begin{equation}
    \label{Eq lambda v B}
   \lambda_B^v  = \sqrt{ \frac{\text{tr}(\mathbf{F}_B^v (\mathbf{F}_B^v)^T}{3}}
   \end{equation}
    \item \( \xi \) is a strain adjustment factor,
    \item \( c \) is the strain exponent governing stretch sensitivity,
    \item \( \tau \) is the magnitude of the deviatoric stress in Network B, given by Eq. (\ref{Eq tau}),
    \item \( \hat{\tau} \) is the shear flow resistance,
    \item \( m \) is the shear flow exponent controlling rate sensitivity.
\end{itemize}
Thus, the viscoelastic parameters, $c$, $\hat{\tau}$, $\xi$, and $m$, along with the elastic parameters $\mu_1$ and $\mu_2$, need to be calibrated with experimental data to use this model to predict the behavior of FAA-3.  
\section{Model Calibration and Validation}
\label{sec:calibration_validation}

The parameters of the parallel network model were calibrated to minimize the error between model predictions and experimental data.  Specifically, experimental data from load-hold-unload tests (Section \ref{sec:Load-Hold-Unload Results}) and some ultimate tensile tests (Section \ref{sec:Ultimate Tensile Results}) were used to calibrate the model from Section \ref{sec:model} for FAA-3.
During the calibration process, values for the parameters listed in Table \ref{tab:ambient_parameters} were determined.  Initially, the parameters were calibrated for only ambient conditions. Subsequently, how the parameters vary with relative humidity and temperature was determined using the experimental data from Sections \ref{sec:Load-Hold-Unload Humidity Results} and \ref{sec:Load-Hold-Unload Temperature Results}, respectively.

The calibrated model was validated by predicting the mechanical response under additional experimental scenarios not used in the calibration process, namely the remaining ultimate tensile tests from Section \ref{sec:Ultimate Tensile Results} and the cyclic loading from  Section \ref{sec:Cyclic Loading Results}.

\subsection{Calibration of the Parallel Network Model}
\label{subsec:calibrate}

  

The software \texttt{MCalibration} was used to calibrate the material parameters \cite{polyumod_manual} by minimizing the error between model predictions and experimental data. 
During the optimization process, the Normalized Mean Absolute Difference (NMAD) error was monitored. The NMAD represents the average absolute deviation between predicted and experimental stress values, normalized by the experimental data range. An NMAD below 10\% was considered acceptable.

At first, the target data was from both the ultimate tensile tests and load-hold-unload tests under ambient conditions (RH = 38\% -- 41\%, \(T = \SI{23}-\SI{27}{\celsius}\)) from Figures \ref{fig:Humidity Ultimate Tensile Test eng} and \ref{fig:Humidity_LHU}, respectively. Both these experiments were calibrated simultaneously.  In this case, first a global search was performed, then a simplex method was used to refine the parameters.

The global search used the Covariance Matrix Adaptation Evolution Strategy (CMA-ES), which is a stochastic, derivative-free optimization algorithm well-suited for problems where the initial parameter guess is far from the optimal solution. CMA-ES performs an evolutionary search by adapting the covariance matrix of a multivariate normal distribution to sample candidate solutions, gradually refining them toward a global minimum. The algorithm was configured with a maximum of 1000 generations, a function accuracy threshold of 0.001, and an initial standard deviation of 0.2.

For the simplex method, the Nelder-Mead algorithm was used, which is a gradient-free method that operates by iteratively adjusting a simplex -- a geometric figure composed of multiple trial solutions -- through reflection, expansion, contraction, and shrinkage to move toward a local minimum of the objective function.  The starting guess for the simplex method was the parameters found through the global search.  

For both optimization methods, a single CPU was used, with a maximum optimization time of \SI{60}{\minute} and a limit of 30{,}000 function evaluations. 

The calibrated model parameters under ambient conditions, are shown in Table~\ref{tab:ambient_parameters}.  The calibrated model showed strong agreement with the experimental data; the NMAD error between the model and the experimental data from the ultimate tensile test at RH = 41\% (Figure~\ref{fig:Humidity Ultimate Tensile Test eng}) was 4.61\%, while the NMAD error for the load-hold-unload tests (Figure~\ref{fig:Humidity_LHU}) was 6.75\%. Considering both datasets together, the combined NMAD was calculated to be 5.33\%, demonstrating the model’s reliable performance across different loading protocols under the same environmental conditions.
\begin{table}[htbp]
\fontsize{9pt}{8pt}\selectfont
\centering
\caption{Parameters in the parallel network model calibrated for ambient conditions (RH = 38 -- 41\%, T = \SI{23}-\SI{27}{\celsius}).}
\label{tab:ambient_parameters}
\begin{adjustbox}{max width=0.95\textwidth}
\begin{tabularx}{\linewidth}{|>{\centering\arraybackslash}X
                            |>{\centering\arraybackslash}X
                            |>{\centering\arraybackslash}X
                            |>{\centering\arraybackslash}X
                            |>{\centering\arraybackslash}X
                            |>{\centering\arraybackslash}X
                            |>{\centering\arraybackslash}X
                            |>{\centering\arraybackslash}X|}
    \hline
    \textbf{$\mu_1$ (MPa)} & \textbf{$\mu_2$ (MPa)} & \textbf{$c$} & \textbf{$\xi$} & \textbf{$\hat{\tau}$ (MPa)} & \textbf{$m$} & \textbf{NMAD (\%) -- UTT} & \textbf{NMAD (\%) -- LHU} \\
    \hline
    \textbf{Shear modulus 1} & \textbf{Shear modulus 2} & \textbf{Strain exponent} & \textbf{Strain adjustment factor} & \textbf{Shear flow resistance} & \textbf{Shear flow exponent} & \textbf{Ultimate tensile test} & \textbf{Load-hold-unload test} \\
    \hline
    16.1 & 249 & -0.7 & 0.078 & 69.9 & 11.28 & 4.61 & 6.75 \\
    \hline
\end{tabularx}
\end{adjustbox}
\end{table}

It is important to note that while both tests used for calibration were in ambient conditions, the relative humidity and temperatures during the tests differed due to environmental changes.  The differences in relative humidity and temperature are not considered during the calibration process, and it is assumed that the model parameters would not be significantly affected by the approximately 3\% difference in relative humidity nor the approximately $\SI{2.7}\celsius$ difference in temperature.

After calibrating the model parameters for ambient conditions, these values were used as the initial guess for calibrating the next closest experiment in terms of relative humidity (RH = 78\%). 
In this case, only the load-hold-unload data at RH = 78\% from Figure~\ref{fig:Humidity_LHU} was used as the target experimental data, and only the simplex method (Nelder-Mead) was employed.  This strategy yielded a good fit, with a normalized mean absolute deviation (NMAD) error of 4.7\%, well below the 10\% threshold. The parameters obtained from the 78\% RH calibration then served as the initial guess to calibrate the model to the 85\% RH data in Figure~\ref{fig:Humidity_LHU}. This sequential calibration approach, where each optimization used as an initial guess the parameters found for the previous humidity level, together with the load-hold-unload data from Figure~\ref{fig:Humidity_LHU} and the simplex method (Nelder-Mead), was applied to all subsequent experiments at various relative humidity levels.  
%
In every case, the NMAD error remained below 10\%. A summary of the calibrated parameters and associated NMAD errors is presented in Table~\ref{tab:parameters_rh}.

Notice that not all parameters are included in Table~\ref{tab:parameters_rh}.  This is because during the calibration process, it was observed that certain parameters, specifically ${\xi}$, $c$, and $m$ showed minimal sensitivity to changes in RH. Thus, to keep the model as simple as possible, these parameters were treated as constants with the values matching those found in ambient conditions, namely, ${\xi} = 0.078$, $c = -0.7$, and $m = 11.28$. In contrast, the remaining parameters---$\mu_1$, $\mu_2$, and $\hat{\tau}$---were found to vary significantly with relative humidity.

After calibration to various relative humidity levels, the model was calibrated to various temperatures using the submerged tests from Figure \ref{fig:Temp_LHU} and the simplex method (Nelder-Mead).  In this case, the parameters calibrated to 95\% RH load-hold-unload data served as the initial guess for the submerged conditions at 28\SI{}{\celsius}. Then the resulting parameters obtained for submerged conditions at 28\SI{}{\celsius} were used as the initial guess for submerged conditions at 40\SI{}{\celsius}.  Again, the parameters ${\xi}$, $c$, and $m$ showed minimal sensitivity to changes in temperature, and therefore, were considered constant with values matching those found for ambient conditions.  The resulting model parameters for submerged conditions at various temperatures are shown in Table~\ref{tab:parameters_t} along with the NMAD error, which is less than 10\% for all cases. 

Note that Table~\ref{tab:parameters_t} shows two sets of parameters for each temperature.  This is because there were two experimental trials at each temperature, and each trial was calibrated separately. By performing these calibrations separately, the potential distribution of material parameters under the same test conditions can be observed.  For both temperatures, the parameters found using the different trials are relatively close.  The maximum difference of any parameter found from the different trials is 7.58\%. 

Also note that the NMAD is generally higher in Table~\ref{tab:parameters_t} than in Table~\ref{tab:parameters_rh}.  
This suggests that submerged conditions will not be predicted as well with the model.  To mitigate these poorer predictions, one might consider performing a global search as part of the calibration under submerged conditions, extending the optimization time, and/or increasing limit number of functional evaluations.  However, as submerged conditions are not expected in MS DAC, this was not attempted in this work, and these higher errors were still considered acceptable.  

\begin{figure}[H]
    \centering
    \includegraphics[width=0.9\textwidth]{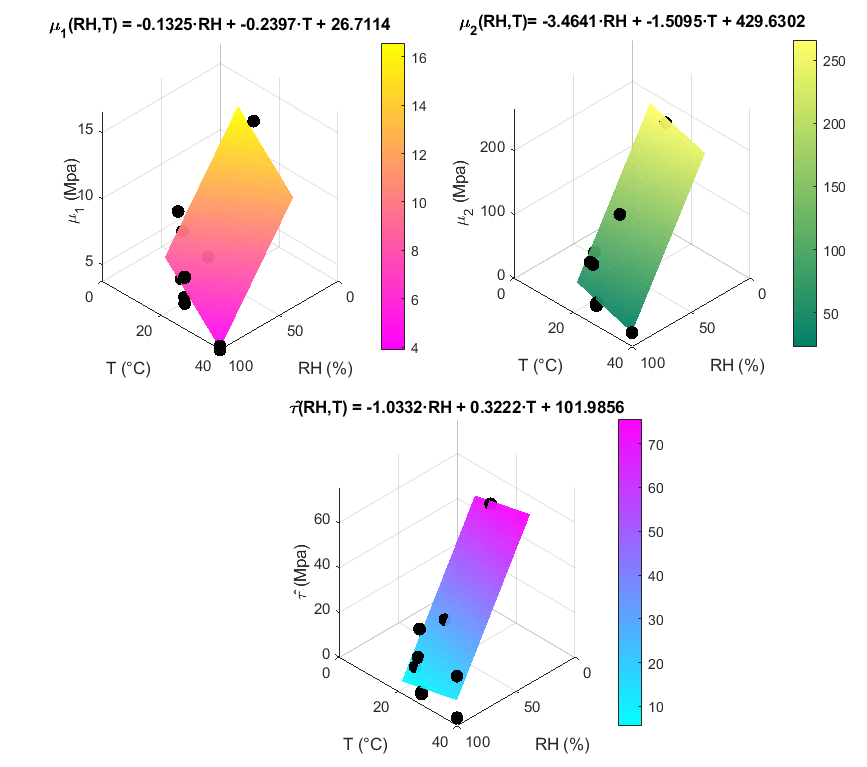} %
    \caption{Material parameters interpolated as a linear function of temperature ($T$) and relative humidity (RH).}
    \label{fig:RHandtemp}
\end{figure}

%
Finally, empirical relationships were developed to express $\mu_1$, $\mu_2$, and $\hat{\tau}$ as linear functions of RH and temperature using the data in Tables~\ref{tab:parameters_rh} and ~\ref{tab:parameters_t}. 
These trends, shown in Figure~\ref{fig:RHandtemp}, enable predictive modeling for untested environmental conditions.

\begin{table}[htbp]
\fontsize{9pt}{9pt}\selectfont
    \centering
    \caption{Calibrated Parameters for Different Relative Humidity (RH) Levels}
    \label{tab:parameters_rh}
    \begin{adjustbox}{max width=0.9\textwidth}
    \begin{tabularx}{\linewidth}{|>{\centering\arraybackslash}X
                                |>{\centering\arraybackslash}X
                                |>{\centering\arraybackslash}X
                                |>{\centering\arraybackslash}X
                                |>{\centering\arraybackslash}X
                                |>{\centering\arraybackslash}X|}
        \hline
        \textbf{RH (\%)} & \textbf{$T$ (\SI{}{\celsius})} & \textbf{$\mu_1 (Mpa)$} & \textbf{$\mu_2 (Mpa)$} & \textbf{$\hat{\tau} (Mpa)$} & \textbf{NMAD} \\
        \hline
        38  & 26.6 & 16.1 & 249 & 69.9 & 5.68 \\
        \hline
        78  & 27.1 & 7.97  & 150 & 30.8 & 4.7  \\
        \hline
        85  & 21.3 & 9.55  & 82.45  & 24.3 & 9.28 \\
        \hline
        93  & 24   & 6.72  & 78.9  & 16.2 & 6.05 \\
        \hline
        95  & 23.8 & 11.9 & 84.3  & 12.4 & 8.89 \\
        \hline
    \end{tabularx}
    \end{adjustbox}
\end{table}

\begin{table}[htbp]
\fontsize{12pt}{9pt}\selectfont
    \centering
    \caption{Calibrated Parameters for different Temperatures at submerged condition  (RH=100\%)}
    \label{tab:parameters_t}
    \begin{adjustbox}{max width=0.7\textwidth}
    \begin{tabularx}{\linewidth}{|>{\centering\arraybackslash}X 
                                |>{\centering\arraybackslash}X 
                                |>{\centering\arraybackslash}X 
                                |>{\centering\arraybackslash}X 
                                |>{\centering\arraybackslash}X 
                                |>{\centering\arraybackslash}X|}
        \hline
        \textbf{$T$ (\SI{}{\celsius})} & \textbf{$\mu_1 (Mpa)$} & \textbf{$\mu_2 (Mpa)$} & \textbf{$\hat{\tau} (Mpa)$} & \textbf{NMAD} \\
        \hline
        28 & 7.73 & 32.3 & 5.07 & 8.89 \\
        \hline
        28 & 7.67  & 32.3 & 5.07 & 8.95  \\
        \hline
        40 & 4.06 & 22.7  & 3.73 & 9.75 \\
        \hline
        40 & 3.77  & 22.3  & 3.79 & 9.73 \\
        \hline
    \end{tabularx}
    \end{adjustbox}
\end{table}

\subsection{Model Validation and Prediction}
\label{subsec:validate}
The predictive capability of the parallel network model was evaluated by simulating experimental tests that were not included in the calibration process. This approach enabled an assessment of the model’s generalizability across different loading scenarios and environmental conditions. Specifically, the model was validated against ultimate tensile tests conducted at various strain rates and relative humidity levels, not including the ambient condition used in calibration, and the cyclic loading tests performed at room temperature under varying humidity levels.

\begin{figure}[H]
    \centering
    \begin{subfigure}[b]{0.48\textwidth}
        \centering
        \includegraphics[width=\textwidth]{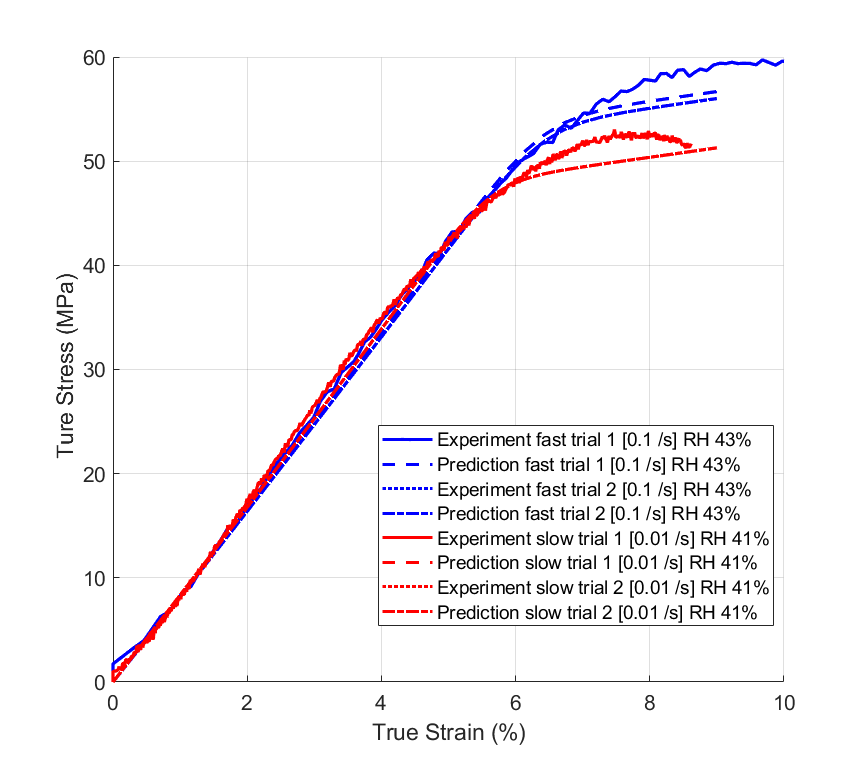}
        \caption{Ultimate tensile tests at different engineering strain rates (0.001 and 0.1 s$^{-1}$) under ambient conditions (RH approximately 42\% and temperature \SI{23}{\celsius}).}
        \label{fig:RHrate}
    \end{subfigure}
    \hfill
    \begin{subfigure}[b]{0.48\textwidth}
        \centering
        \includegraphics[width=\textwidth]{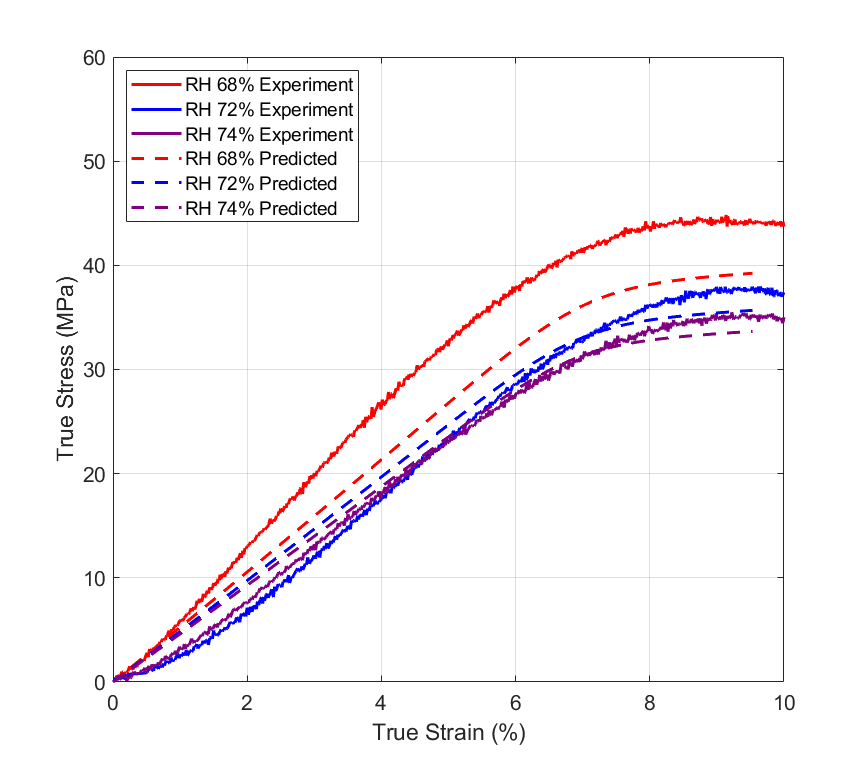}
        \caption{Ultimate tensile tests at a fixed engineering strain rate (0.001 s$^{-1}$) with varying humidity and ambient temperature (\SIrange{22}{23}{\celsius}
).}
        \label{fig:RHhumid}
    \end{subfigure}
    \caption{Comparison of predicted and experimental true stress-true strain responses for the ultimate tensile tests.}
    \label{fig:combined_RH_plots}
\end{figure}

For the ultimate tensile tests, 
FAA-3 specimens were stretched to failure under both slow (0.001 s$^{-1}$) and fast (0.1 s$^{-1}$) strain rates at \SI{23}{\celsius} temperature and 42\% relative humidity, as shown in Figure~\ref{fig:RHrate}. To examine environmental sensitivity, additional simulations were performed at the same slow strain rate (0.001 s$^{-1}$), but under higher humidity levels (RH 68\%,  72\%, and 74\%), still at room temperature (\SIrange{22}{23}{\celsius}). Predictions of these tests, along with the experimental data, are shown in Figure~\ref{fig:RHhumid}. 
The parameters used in these simulations use the environmental dependencies shown in Figure~\ref{fig:RHandtemp} as well as the fixed values for $\xi$, $c$, and $m$ from Table~\ref{tab:ambient_parameters}. Because the model does not include a failure criterion, it cannot capture the stress-strain behavior beyond the point of maximum stress. In the experimental data, the peak stress typically occurs at an engineering strain of approximately $10\%$. Therefore, the model predictions and experimental data in Figure~\ref{fig:combined_RH_plots} are truncated around this point to 
avoid attempting to model a behavior that falls outside the scope of the current constitutive framework.

As illustrated in Figure~\ref{fig:combined_RH_plots}, the predicted stress--strain responses closely matched the experimental data, accurately capturing both 
the linear and non-linear regions.  
The model does particularly well in predicting the data in Figure~\ref{fig:RHrate}, as those conditions closely mirror the ultimate tensile tests used in the calibration process, and those tests had some of the lowest NMAD error of all the calibration tests.  
Interestingly, the model also performs quite well in predicting the results at 72\% and 74\% RH, but less well in predicting the results at 68\% RH.  This is likely because there was no experimental data used for calibration between approximately 41\% RH and 78\% RH.  Therefore,  72\% and 74\% RH were relatively close to the data used in calibration, while 68\% RH was less so.  
This highlights some of the challenges in predicting experimental results under various environmental conditions and may indicate that a more sophisticated model for how parameters vary with humidity might be necessary.  

\begin{figure}[H]
    \centering

    \begin{subfigure}[b]{0.5\textwidth}
        \centering
        \includegraphics[width=\textwidth]{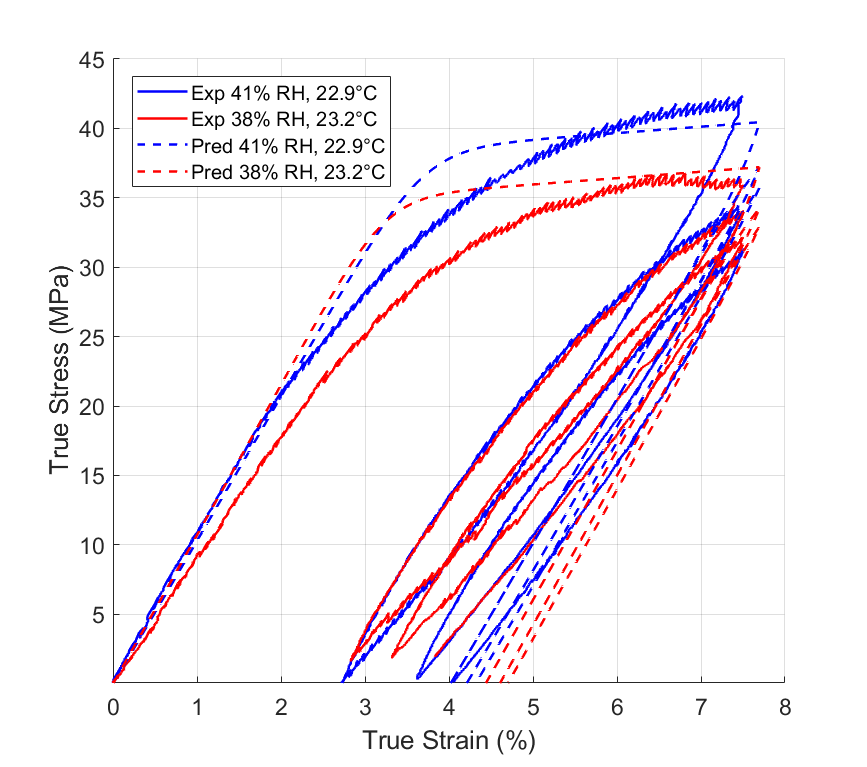}
        \caption{True stress–true strain response.}
        \label{fig:cyclic_a}
    \end{subfigure}
    \hfill
    \begin{subfigure}[b]{0.5\textwidth}
        \centering
        \includegraphics[width=\textwidth]{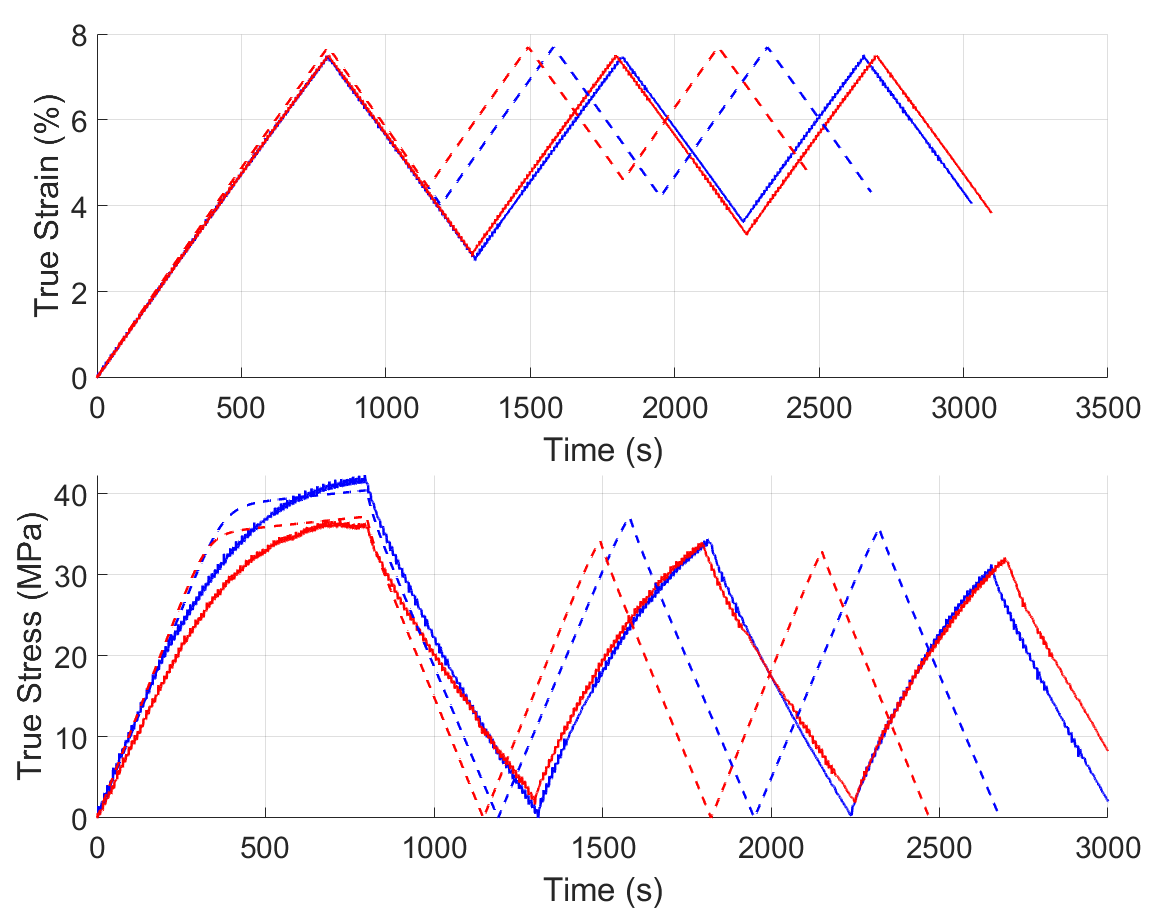}
        \caption{True strain–time and true stress–time responses.}
        \label{fig:cyclic_b}
    \end{subfigure}

    \caption{Comparison of experimental and predicted cyclic loading behavior at ambient conditions (RH = 38--41\%, $T$ = \SI{27}{\celsius}). (a) True stress–strain response. (b) True strain and true stress as functions of time.}
    \label{fig:cyclicprediction}
\end{figure}

For cyclic loading, the model simulated the response of FAA-3 under deformation to 8\% engineering strain at a strain rate of 0.001~s$^{-1}$, followed by unloading to zero force at a rate of $-0.001$~s$^{-1}$, repeated for three cycles. These simulations used the calibrated parameters as functions of temperature/humidity as shown in Figure~\ref{fig:RHandtemp}, along with the fixed values for $\xi$, $c$, and $m$ from Table~\ref{tab:ambient_parameters}. The predicted cyclic response is compared to the experimental data in Figure~\ref{fig:cyclicprediction}, and shows strong agreement overall. 

As expected, because the model does not perfectly capture the effects of plastic deformation, the discrepancy between the experimental and simulated responses becomes increasingly pronounced with each cycle. This accumulation of error reflects the model’s current limitations in representing residual deformations under repeated loading and unloading conditions.

\section{Conclusion}
The mechanical behavior of FAA-3 was experimentally explored to quantify the tensile, cyclic, and time-dependent mechanical response of FAA-3, including humidity and temperature effects. The experiments performed were DMA, hygroscopic swell, thermal expansion, ultimate tensile, load-hold-unload, and cyclic loading tests. The DMA test 
showed non-negligible viscous effects 
throughout the test and that the glass transition temperature does not occur within the range of \SI{-150}{\celsius} to \SI{200}{\celsius}, but may occur around \SIrange{200}{250}{\celsius}.  Thus, glass transition is not expected for  dry FAA-3 used in direct air capture. 
The hygroscopic swell results demonstrated that the hygroscopic swelling strain in the in-plane and thickness directions were similar and displayed an approximately linear trend with change in water content. The swell coefficient ($\bm{\beta}$) was found to be in the range of 1.02 $\times$ 10$^{-5}$ m$^3$/mol in the in-plane direction and 9.24 $\times$ 10$^{-6}$ m$^3$/mol in the thickness direction, suggesting nearly isotropic behavior. 
The swell results were also used to calculate the excess volume of mixing and indicate that FAA-3 exhibits greater volume expansion than expected from ideal mixing, implying that water uptake not only fills available free volume but also induces structural relaxation and polymer chain expansion. 
Thermal expansion tests showed an approximately linear relationship between thermal strain and change in temperature with a coefficient of thermal expansion ($\alpha$) of $5.07 \times 10^{-3}$/\SI{}{\celsius}. The ultimate tensile test provided an initial understanding of the material's behavior and properties. The elongation at break was around 13\% mechanical strain for a dry sample and did not change much when exposed to moisture. The ultimate tensile strength of FAA-3 is approximately 50 MPa in a dry state, and that value decreases with moisture content. A small decrease in strength with an increase in the rate at which the sample is loaded was also observed, but this influence is small for the range of rates used in this work. 

The load-hold-unload tests were conducted with at various relative humidities and temperatures to observe stress relaxation during the hold and residual strains during the unload.  
 Through these tests, it was found that the load needed to pull FAA-3 to a certain strain decreases as the humidity or temperature increases, the amount of stress relaxation that occurs increases with strain but decreases with increasing humidity or temperature, and the plastic deformation increases with increasing humidity or temperature. 
These changes are generally as expected, as humidity increases lubrication and therefore decreases stiffness and increases plastic deformation, and temperature increases molecular vibrations, again making deformation and slippage easier.  
It was also found that polymers may have different maximum water uptake in the liquid and saturated vapor phases, as implied by the observation that a 5\% change in RH from 95\% to full submersion led to an approximately 2.5-fold difference in stiffness.


With these experimental findings for the mechanical behavior of FAA-3, we determined that the model needed to include viscous effects and residual strains upon unloading to capture the experimentally observed stress-strain response under uniaxial tension and cyclic loading conditions. 
Specifically, a two-element parallel-network framework was adopted, utilizing Neo-Hookean hyperelastic springs for elastic responses and an adapted Bergstr\"{o}m-Boyce viscous dashpot for time-dependent behavior. This model was calibrated using experimental data from ultimate tensile and load-hold-unload tests across various humidity and temperature conditions, achieving Normalized Mean Absolute Difference (NMAD) errors below 10~\% for all data used in calibration. 
Empirical relationships for parameters related to elastic deformation, \(\mu_1\) and \(\mu_2\), and shear flow, \(\hat{\tau}\), allowed the model to incorporate changes in mechanical behavior with relative humidity and temperature.  With these empirical relationships, the model can predict the response of FAA-3 in untested environmental scenarios. The model was validated through simulations of ultimate tensile tests and cyclic loading tests, which were not used in calibration.  The results shown in Figures~\ref{fig:combined_RH_plots} and \ref{fig:cyclicprediction} suggest that the model can adequately capture FAA-3’s mechanical response.  However, as expected, cyclic loading is more difficult to predict as errors in each cycle accumulate.   

Based on the findings of this work, and the presumed loading of FAA-3 during MS DAC (see Figure \ref{fig:schematic}), it will be critical that this model captures how humidity affects the mechanical behavior, and less emphasis may be placed on the role of temperature on the material properties. This is because humidity variations will be the driver for the use of FAA-3 in MS DAC.  While the system is in use, significant RH gradients will be present.  During start-up or shut-down, the swelling of the internal portion of the hollow fibers and/or the contraction of the outer portion of the hollow fibers may cause internal cracking and material failure.  Temperature variations are primarily expected due to external environmental changes.  Nonetheless, significant temperature gradients may exist, and thus, the temperature dependence of the elastic and some viscous parameters may be important. 

Throughout this work, emphasis was placed on the in-plane behavior of FAA-3, with the notable exception of the moisture swelling through the thickness of the sample being measured and modeled.  This is because,  during MS, FAA-3 is most likely to be in bending (see Figure \ref{fig:schematic}).  The forces from the airflow and the relative humidity gradients all cause bending of the hollow fibers.  Nonetheless, the model presented is fully 3-dimensional and can be applied to any load conditions of FAA-3.  However, the model has only been validated against 1-dimensional behavior, and this limit must be considered when applying this model to the behavior of FAA-3.  

Future work aims to integrate the mechanical model from this paper with chemical transport to effectively predict the chemo-mechanical behavior of FAA-3 during the carbon capture process. With this coupled approach, we can study how the mechanical behavior is affected by the transport processes and vice versa, simulate the material's response within the MS DAC system more accurately, and optimize the design of MS DAC systems.

\section*{Acknowledgments}
This material is based upon work supported by the U.S. Department of Energy, Office of Science, Office of Basic Energy Sciences under Award Number DE-SC0023343.
The authors would also like to acknowledge Dr. Constantin Ciocanel and the Multifunctional Materials and Adaptive Systems Lab at Northern Arizona University,   Dr. Timothy Becker and the Bioengineering Devices Lab at Northern Arizona University. 

\section*{Declaration of generative AI and AI-assisted technologies in the writing process.}
During the preparation of this work the authors used Chat GPT in order to improve the readability and language of the manuscript. After using this tool, the authors reviewed and edited the content as needed and assumed full responsibility for the content of the published article.
 \bibliographystyle{elsarticle-num} 
 \bibliography{references}





\end{document}